\documentstyle[prl,aps,epsf]{revtex}
\begin{document}
\newcommand {\lscof} {\rm La_{1.85}Sr_{0.15}CuO_4}
\newcommand {\lscos} {\rm La_{1.86}Sr_{0.14}CuO_4}
\newcommand {\pipi} {(\pi/a,\pi/a)}
\newcommand {\ybco} {\rm YBa_2Cu_3O_{6+x}}
\newcommand {\ybcos} {\rm YBa_2Cu_3O_7}
\newcommand {\ybcoss} {\rm YBa_2Cu_3O_{6.96}}
\newcommand {\lsco} {\rm La_{2-x}Sr_xCuO_4}
\newcommand {\bsco} {\rm Bi_2Sr_2CaCu_2O_8}
\newcommand {\bscoo} {\rm Bi_2Sr_2CuO_x}
\newcommand {\bscot} {\rm Bi_2Sr_2CaCu_2O_x}
\newcommand {\ybce} {\rm YBa_2Cu_4O_8}
\newcommand {\ybcox} {\rm YBa_2Cu_3O_{6.63}}
\newcommand {\Tc} {T_c}
\newcommand {\beq} {\begin{equation}}
\newcommand {\eeq} {\end{equation}}
\newcommand {\beqa} {\begin{eqnarray}}
\newcommand {\eeqa} {\end{eqnarray}}
\newcommand {\cu} {\rm ^{63}Cu}
\newcommand {\ox} {\rm ^{17}O}
\newcommand {\tcu} {(^{63}T_1T_{c})^{-1}}
\newcommand {\tcup} {(^{63}T_1T_{ab})^{-1}}
\newcommand {\too} {(^{17}T_1T)^{-1}}
\newcommand {\ra} {^{63}R}
\newcommand {\bp}  {{\bf p}}
\newcommand {\bq}  {{\bf q}}
\newcommand {\bk}  {{\bf k}}
\newcommand {\bQ}  {{\bf Q}}
\title{NMR and Neutron Scattering Experiments on the Cuprate Superconductors:
A Critical Re-Examination}
\author{Y. Zha$^{(a)}$, V. Barzykin$^{(b)}$ and D. Pines$^{(a)}$}
\address{$^{(a)}$
Department of Physics and Science and Technology Center for
Superconductivity, University of Illinois at Urbana-Champaign,
1110 West Green Street, Urbana, IL 61801}
\address{$^{(b)}$Department of Physics, University of British Columbia,
6224 Agricultural Rd., \\ Vancouver, B. C. V6T 1Z1, Canada}
\preprint{}
\date{\today}
\maketitle

\begin{abstract}
We show that it is possible to 
reconcile NMR and neutron scattering experiments on both
$\lsco$ and $\ybco$, by making use of the Millis-Monien-Pines mean field 
phenomenological expression for the dynamic spin-spin response function,
 and reexamining the standard Shastry-Mila-Rice hyperfine
Hamiltonian for NMR experiments. The recent neutron scattering 
results of Aeppli 
$et~al$\cite{aeppli} on $\lscos$ are shown to agree quantitatively 
with the NMR measurements of
$^{63}T_1$ and the magnetic scaling behavior proposed by Barzykin and 
Pines.\cite{barzykin} The reconciliation of the  $\ox$ relaxation
rates with the degree of incommensuration in the spin fluctuation 
spectrum seen in neutron experiments 
is achieved by introducing a new transferred hyperfine coupling
$C'$ between $\ox$ nuclei and their next nearest neighbor $Cu^{2+}$ spins;
this leads to a near-perfect cancellation of the influence of the 
incommensurate spin
fluctuation peaks on the  $\ox$ relaxation rates of $\lsco$. The inclusion of the
new $C'$ term also leads to a natural explanation, within the one-component model,  the different temperature
dependence of the anisotropic $\ox$ relaxation rates for different field
orientations, recently observed by Martindale $et~al$.\cite{martindale}  The measured
significant decrease with doping of
the anisotropy ratio, $\ra= ^{63}T_{1ab}/^{63}T_{1c}$ in $\lsco$ system, from
 $\ra =3.9$ for ${\rm La_2CuO_4}$ to $\ra\simeq3.0$ for $\lscof$ is
made compatible with the doping dependence of the shift in the incommensurate
spin fluctuation peaks measured in neutron experiments, by suitable choices of
the direct and transferred hyperfine coupling constants $A_{\beta}$ and B.
\end{abstract}
\section{introduction}
The magnetic behavior of the planar excitations in the cuprate superconductors
continues to be of central concern to the high temperature superconductivity
community.  Not only does it provide significant constraints on candidate
theoretical descriptions of their anomalous normal state behavior, but it may
also hold the key to the physical origin of high temperature
superconductivity.  Recently two of us have used the results of NMR experiments
to determine the magnetic phase diagram for the $La_{2-x}Sr_xCuO_4$ and
$YBa_2Cu_3O_{6+x}$ systems.\cite{barzykin}  We found that for both systems bulk
properties, such as the spin susceptibility, and probes in the vicinity of the
commensurate antiferromagnetic wave vector ($\pi,\pi$), such as $^{63}T_1$, the
$^{63}Cu$ spin relaxation time, and $^{63}T_{2G}$, the spin-echo decay time,
display $z=1$ scaling and spin pseudogap behavior over a wide regime of
temperatures.  On the other hand, the neutron scattering experimental results
of Aeppli {\it et al}\cite{aeppli} on $La_{1.86}Sr_{0.14}CuO_4$ which probe
directly $\chi''(q,\omega)$, the imaginary part of the spin-spin response
function, while  supporting this proposed scaling behavior,
 at first sight appear incapable of explaining NMR experiments on this
system.

This apparent contradiction between the results of NMR and neutron scattering
experiments, both of which probe $\chi(q,\omega)$ in $La_{1.86}Sr_{0.14}CuO_4$,
is but one of a series of such apparent contradictions.  For example, in the
$YBa_2Cu_3O_{6+x}$ system, NMR experiments on $^{63}Cu$ and $^{17}O$ nuclei in
both $YBa_2Cu_3O_7$\cite{hammel1} and $YBa_2Cu_3O_{6.63}$\cite{takigawa} require
the presence of strong antiferromagnetic correlations between the planar
$Cu^{2+}$ spins, and a simple mean field description of the spin-spin response
function  with a temperature dependent magnetic correlation length
$\xi \agt 2$, was shown to provide a quantitative description of the measured
results for $^{63}T$ and $^{17}T_1$ in $YBa_2Cu_3O_7$,\cite{mmp1} and
$YBa_2Cu_3O_{6.63}$.\cite{monien}  Yet neutron scattering experiments on
$YBa_2Cu_3O_7$\cite{bourges,mook,keimer} and $YBa_2Cu_3O_{6.63}$,\cite{tranquada} find only
comparatively broad, temperature-independent, peaks in $\chi''(q,\omega)$,
corresponding to a quite short ($\xi \alt 1$) temperature-independent magnetic
correlation length.  The apparent contradiction is especially severe for the
$\lsco$ system, where neutron scattering experiments show at low temperatures
four incommensurate peaks in the spin fluctuation spectrum, whose position
depends on the level of Sr doping,\cite{aeppli2} while the
quantitative explanation (using the same one-component phenomenological
description which worked for the $\ybco$ system) of the measurements of
$^{63}T_1$ and $^{17}T_1$ in this
system requires that the spin fluctuations be peaked at ($\pi,\pi$), or nearly
so.\cite{monthoux,thelen}  Viewed from the NMR perspective, there are two major problems
with four incommensurate spin fluctuation peaks.  First, the Shastry-Mila-Rice
(SMR) form factor,\cite{shastry,mila} which, provided the peaks are nearly at
($\pi,\pi$), effectively screens neighboring 
$^{17}O$
nuclei from the presence of the strong peaks in the nearly
localized $Cu^{2+}$ spin spectrum required to explain the anomalous
temperature-dependence behavior of $^{63}T_1$,  fails to do so for the considerable degree of incommensuration
in the peaks at ($\pi,[\pi \pm \delta])$ and  $([\pi \pm \delta], \pi$) seen in
$La_{1.86}Sr_{0.14}CuO_4$.\cite{thelen,zha,walstedt}  As a result $^{17}T_1$ picks up a substantial
anomalous temperature dependence which is not seen
experimentally.  Second, with the doping-independent values of the hyperfine
couplings which appear in the SMR form factors for a commensurate spectrum,
the calculated anisotropy of $^{63}T_1$ for the incommensurate peaks seen by
neutrons is in sharp variance with what is seen in the NMR experiments.\cite{thelen}

Two ways out of these apparent contradictions have been proposed.  One view
is that the spin fluctuation peaks seen in the neutron scattering experiments
reflect the appearance of discommensuration, not incommensuration; on this
view, the $\lsco$ system contains domains in which the spin fluctuation peaks are
commensurate (so that there are no problems with $^{17}T_1$), but what neutrons,
a global probe, see is the periodic array of the domain walls.\cite{cps}
A second view is that a one-component description of
$\chi(\bq,\omega)$ is not feasible; rather, the transferred hyperfine
coupling between the nearly localized $Cu^{2+}$ spins and the $^{17}O$ nuclei
is presumed to be very weak, and the $^{17}O$ nuclei are assumed to be relaxed by a different mechanism,
whence the nearly Korringa-like behavior of $^{17}T_1$.\cite{walstedt}
  A further challenge to
a one-component description has come from the very recent work of Martindale
{\it et al},\cite{martindale} who find that their results for the
temperature-dependence of the planar anisotropy of $^{17}T_{1\alpha}$ for 
different field orientations appear
incompatible with a one-component description.

In the present paper we present a third view:  that the one-component
phenomenological description is valid, but what requires modification are the
hyperfine couplings which appear in the SMR Hamiltonian which describes
planar nuclei coupled to nearly localized $Cu^{2+}$ spins.  We find that by
introducing a transferred hyperfine coupling $C^\prime$, 
between the next nearest neighbor
$Cu^{2+}$ spins and a $^{17}O$ nucleus, the nearly antiferromagnetic part of
the strong signals emanating from the $Cu^{2+}$ spins can be far more
effectively screened than is possible with only a nearest neighbor transferred
hyperfine coupling, so that the existence of four incommensurate peaks in the
$\lsco$ system can be made compatible with the $^{17}T_1$ results.  We also find
that by permitting the transferred hyperfine coupling, $B$, between a $Cu^{2+}$
spin and its nearest neighbor $^{63}Cu$ nucleus to vary with doping, we can
explain the trend with doping of the anisotropy of $^{63}T_1$ in this system.
We then use these revised hyperfine couplings to reexamine the extent to which
the recent results of Aeppli {\it et al}\cite{aeppli} on $\lscos$ can be explained
quantitatively by combining the Millis-Monien-Pines (hereafter MMP) response function\cite{mmp1}
 with the scaling arguments put forth by Barzykin and
Pines.\cite{barzykin}  We find that they can, and are thus able to reconcile 
the neutron
scattering and NMR experiments on this member of the $\lsco$ system.

We present as well the results of a reexamination of the NMR and neutron
results for the $\ybco$ system.  Here we begin by making the ansatz that it is
the presence of incompletely resolved incommensurate peaks which is
responsible for the broad lines seen in neutron experiments.  We follow
Dai {\it et al},\cite{mook} who suggest the increased line width for $\ybcos$
seen along the zone diagonal directions  reflects the presence of four
incommensurate peaks, located at $\bQ_i = (\pi \pm \delta, \pi \pm
\delta)$, a proposal which is consistent with the earlier measurements of Tranquada $et~al$
for $YBa_2Cu_3O_{6.6}$ \cite{tranquada}.  We then find that  incommensuration
 can be made compatible with NMR
experimental results provided the transferred hyperfine coupling constant,
$B$, is doping dependent in this system as well.  Moreover, on considering $^{17}T_1$ for $\ybcos$, we find that the anomalous temperature dependence of the 
planar anisotropy of $^{17}T_1$ measured by Martindale $et~al$\cite{martindale}
constitutes a proof of the validity of our modified one-component model. Thus
incommensuration combined with the presence of the next
nearest neighbor coupling, $C'$, leads to results  which are consistent with 
experiment, and we are able to preserve the
one-component description of the planar spin excitation spectrum.

The outline of our paper is as follows:  In Section II we review the SMR
description of coupled $Cu^{2+}$ spins and nuclei as well as the mean field
description of $\chi(\bq,\omega)$, and examine the modifications brought
about by incommensuration and next nearest neighbor coupling between $Cu^{2+}$ spins and a
 $^{17}O$ nucleus.  In Section III we review the experimental constraints on the
hyperfine coupling parameters, and present our results for their variation with
doping in both the $\lsco$ and $\ybco$ systems.  We show in Section IV how the $^{63}Cu$ NMR
results can be reconciled with neutron scattering results on $\lscos$, while
in Section V we present a quantitative fit to the $^{17}T_{1c}$ results for the
$La_{1.85}Sr_{0.15}CuO_4$ based on the four incommensurate peaks in the spin
fluctuation spectrum expected from neutron scattering.  We show in Section VI
how the anomalous results of Martindale {\it et al}\cite{martindale} for 
the $\ybco$ system can be
explained using our modified one-component model, and in Section VII we
present our conclusions.

\section{A Generalized Shastry-Mila-Rice Hamiltonian}
On introducing a hyperfine coupling $C^\prime_{\alpha,\beta}$  between the the $^{17}$O  nuclei and
their next nearest neighbor $Cu^{2+}$ spins, we can rewrite the SMR hyperfine
Hamiltonian for the $\cu$ and $\ox$ nuclei as:
\beqa
H_{hf}=& &^{63}I_{\alpha}({\bf r}_i) \Big[ \sum_{\beta} A_{\alpha,\beta}
S_{\beta}({\bf r}_i)+B
\sum_{j}^{nn}S_{\alpha}({\bf r}_j)\Big] \nonumber \\
&+& ^{17}I_{\alpha}({\bf
r}_i)\Big[C_{\alpha,\beta}\sum_{j,\beta}^{nn}S_{\beta}(
{\bf r}_j) + C^{\prime}_{\alpha,\beta}
\sum_{j,\beta}^{nnn}S_{\beta}(
{\bf r}_j)\Big] 
\label{newh}
\eeqa
where $A_{\alpha,\beta}$ is the tensor for the direct, on-site coupling of
the ${\rm ^{63}Cu}$ nuclei to the Cu$^{2+}$ spins, $B$ is the strength of the
transferred hyperfine coupling of the ${\rm ^{63}Cu}$ nuclear spin to the
four nearest neighbor Cu$^{2+}$
spins, $C_{\alpha,\beta}$ is the transferred hyperfine coupling of the $\ox$ nuclear spin to its nearest neighbor Cu$^{2+}$ spins, and $C'_{\alpha,\beta}$ its coupling to the next nearest neighbor Cu$^{2+}$ spins. 
 The indices ``$nn$'' represent nearest neighbor
electron spins to
the specific nuclei, ``$nnn$'' the next nearest neighbor  $Cu^{2+}$ spins.
As we shall see
below, inclusion of  the  $C^\prime_{\alpha,\beta}$ term enhances the
 cancellation of the anomalous antiferromagnetic spin fluctuations seen
by the $\ox$ nucleus,
 and therefore reduces the
leakage from  incommensurate spin fluctuation peaks to the $\ox$ relaxation
rates. It thus enable us to reconcile the measured  $\ox$
relaxation
rates with the neutron scattering experiments for both $\lsco$ and $\ybco$.

The spin contribution to the NMR Knight shift for the various nuclei
are:\cite{mmp1}

\beqa
^{63}K_c&=&\frac{(A_{c}+4B)\chi_0}{^{63}\gamma_n\gamma_e
\hbar^2} \nonumber \\
^{63}K_{ab}&=&\frac{(A_{ab}+4B)\chi_0}{^{63}\gamma_n\gamma_e
\hbar^2} \nonumber \\
^{17}K_{\beta}&=&\frac{2(C_{\beta}+2C^\prime_{\beta})\chi_0}{^{17}
\gamma_n\gamma
_e\hbar^2}
\eeqa
Here, we have incorporated the new $C^\prime_{\beta}$ term into the $\ox$ 
Knight shift
expression for $^{17}K_\beta$, while the others remain their standard form as
in Ref.\cite{mmp1},  $\gamma_n$ are various nuclei gyromagnetic ratios,
$\gamma_e$ is the electron gyromagnetic ratio,   and $\chi_0$ the static spin
susceptibility.   The indices $c$ and $ab$   refer to the direction
of the applied static magnetic field along the $c$-axis and the $ab$-plane.   
The
spin-lattice relaxation rate, $(^\alpha T_1)^{-1}_\beta$, for nuclei $\alpha$
responding to a magnetic   field in the $\beta$ direction, is:

\beq
^\alpha T^{-1}_{1\beta} =\frac{k_BT}{2\mu_B^2 \hbar^2\omega }
\sum_{\bq} {^\alpha F_\beta(\bq)}\chi^{\prime \prime} (\bq, \omega \rightarrow 0)
\label{t1}
\eeq
where the modified SMR form factors, $^\alpha F_\beta(\bq)$, are now
given by:
\beqa
^{63}F_{c}&=&[A_{ab}+2B(\cos q_xa +\cos q_ya )]^2  \nonumber \\
^{63}F_{ab}&=&\frac{1}{2}[~ ^{63}F_{c} + ^{63}F_{ab}^{eff}] \nonumber \\
^{63}F_{ab}^{eff}&=&[A_c+2B(\cos q_xa + \cos q_ya)]^2     \nonumber \\
^{17}F_\alpha& =& 2\sum_{\alpha_i= \alpha^{\prime}, \alpha^{\prime\prime}}
  \cos^2 \frac{q_xa}{2}\big(C_{\alpha_i} + 2 C^{\prime}_{\alpha_i}
\cos q_ya\big)^2,
\label{form}
\eeqa
Here, $\alpha^{\prime}$ and
$\alpha^{\prime\prime}$ are the directions perpendicular to $\alpha$. The form
factor $^{63}F_{ab}^{eff}$ is the filter for the ${\rm ^{63}Cu}$   spin-echo decay
time $^{63}T_{2G}$\cite{Thelen:Pines}:

\beq
^{63}T_{2 G}^{-2} = \frac{.69}{128\hbar^2\mu_B^4} 
\left\{\frac{1}{N}\, \sum_{\bf q}
F_{ab}^{eff}({\bf q})^2 [\chi'({\bf q},0)]^2   
  - 
\left[\frac{1}{N}\sum_{\bf q} F_{ab}^{eff}({\bf q}) \chi'({\bf q},0) \right]^2
\right\}
\label{t2}
\eeq

The values of the hyperfine constant $C_\alpha$ and $C^\prime_\alpha$ can 
be determined
by the various $\ox$ Knight shift data. In fact, we may obtain these
new values from the ``old'' values of the  hyperfine coupling constant,  
$C_\alpha^{\rm old}$,
which have been well established from fitting the Knight shift
data.\cite{barzykin}
Note we use $C_\alpha$ to represent the new nearest neighbor hyperfine coupling
constant, while the old hyperfine coupling constant is written explicitly
as $C_\alpha^{\rm old}$ throughout the paper.
In order not to change the Knight shift
result of the previous analysis\cite{barzykin}, the new hyperfine
coupling constants should satisfy the following requirement:

\beq
C_{\alpha}+2C^{\prime}_\alpha=C^{\rm old}_{\alpha}=\zeta_\alpha C_c^{\rm old}
\eeq
where
$\zeta_\alpha=C_\alpha^{\rm old}/ C_c^{\rm old}$, and $c$ denotes the 
case of a magnetic field along the $c$-axis.  For $\ybco$, from the previous
analysis of Yoshinari\cite{yoshinari} and Martindale
$et~al$\cite{martindale}, we have  for a field
parallel to the Cu-O bond,  $\zeta_\parallel$=1.42, and  $\zeta_\perp=0.91$ 
for a field perpendicular to
the Cu-O bond direction, while $\zeta_c=1$.  On introducing $r_{\alpha}
\equiv C^{\prime}_\alpha/C_c$ we obtain:

\beqa
C_\alpha&=&C_c^{\rm old}(\zeta_\alpha-\frac{2r_\alpha}{2r_c+1}) \nonumber \\
C^\prime_\alpha&=& C_c^{\rm old}\frac{r_\alpha}{2r_c+1}
\eeqa
Substituting these values of $C_{\alpha}$ and $C^\prime_\alpha$ into
Eq.(\ref{form}), we  obtain the new $\ox$ form factor in terms of 
 $C^{\rm old}_\alpha$:

\beq
^{17}F_\alpha=\frac{2(C_c^{\rm old})^2}{(1+2r_c)^2}\sum_{\alpha_i=
\alpha^{\prime},
\alpha^{\prime\prime}} \cos^2 \frac{q_xa}{2}[\zeta_{\alpha_i}(1+2r_c)
-2r_{\alpha_i}+2r_{\alpha_i} \cos q_ya]^2
\label{nff}
\eeq

Although $C'_\alpha$ may well be anisotropic (as $C_\alpha$ is), in the 
absence of  detailed quantum chemistry calculations, (which lie beyond the purview 
of the present paper)  we
assume  $C'_\alpha$ to be isotropic for illustrative purposes, in which case
 $r_\perp=r_\parallel=r_c \equiv r \equiv C^{\prime}/C_c$.
In Fig.1, we compare our modified form factor
$^{17}F_c $,
 Eq(\ref{nff}), with the standard SMR  form.
It is seen that with a comparatively
small amount of next nearest neighbor coupling, corresponding to
$r \equiv C^{\prime}/C_c=0.25$, the new form factor  is reduced
significantly near
$(\pi/a,\pi/a)$, and is some 30\% narrower near $\bq=0$. This
 indicates that  the
oxygen $(^{17}T_1T)^{-1}$ is less likely to pick up the anomalous
antiferromagnetic contribution near $(\pi/a,\pi/a)$, even when the anomalous
spin fluctuation is slightly spread away from $(\pi/a,\pi/a)$.

We adopt the phenomenological MMP expression for
the spin-spin correlation
function, modified to take into account the presence of four incommensurate
peaks at $\bQ_i$ near $\pipi$\cite{barzykin},

\beq
\chi(\bq,\omega) = \frac{1}{4}\sum_i \frac{\alpha \xi^2 \mu_B^2}
{1+(\bq -\bQ_i)^2
\xi^2 -i \omega/
\omega_{SF}}+ \frac{{\chi}_0(T)}{1-i\pi\omega/\Gamma}
\label{mmp}
\eeq
Here the first term, often called $\chi_{AF}$, represents the anomalous 
contribution
to the spin spectrum, brought about by the close approach to 
antiferromagnetism of the Fermi liquid 
in the vicinity of the  peaks at
$\bq=\bQ_i$ determined by neutron scattering
experiments\cite{aeppli,birgeneau}. For
$\lscos$, $\bQ_i=(\pi/a, [\pi \pm \delta] /a), ([\pi\pm \delta]/a,\pi/a)$,
with $\delta = 0.245\pi$. In Eq.(\ref{mmp}), $\omega_{SF}$ is the 
characteristic frequency
of the spin fluctuations,
$\xi$ is the correlation length, and $\alpha$ is the  scale factor (in units of
states/eV, where $\mu_B$ is the Bohr magneton), 
which relates $\chi_{\bQ_i}$ 
to $\xi^2$; thus the height of each of the four peaks is,
\beq
\chi_{\bQ_i}=\frac{\alpha}{4}\xi^2 \mu_B^2.
\label{eq10}
\eeq
The second term on the right-hand side of Eq.(\ref{mmp}), usually called
$\chi_{FL}$,
is a parameterized form of the normal Fermi Liquid contribution, which is
wave-vector independent over  most of the Brillouin zone; $\Gamma$ is of order 
the Fermi energy. The static  bulk
susceptibility $ \chi_0$, which is generally temperature dependent, has been 
determined for $\lsco$ and $\ybco$
from  copper and
oxygen Knight shift experiments.\cite{barzykin}
For a system with any appreciable antiferromagnetic correlations
($\xi \agt a$), the normal Fermi liquid contribution is small compared to 
$\chi_{AF}$
 for wave vectors in the
vicinity of $\bQ_i$, and plays a negligible role in determining $\tcu$; 
however, because of the filtering action of $^{17}F_{\alpha}$, it makes 
a significant contribution to $\too$. Note that because the
MMP expression for $\chi_{AF}$ is a good
approximation only for wave vectors in the vicinity of the antiferromagnetic 
wave vector
$\bQ_i$, the above expression should not be used in calculating long 
wavelength properties, such as the Knight shift of $\ox$.


For the frequently encountered case of long correlation lengths ($\xi \agt 2a$),
in calculating the various $\cu$
relaxation rates one can approximate $\chi^{\prime\prime}(\bq,\omega)$ by 
$\chi^{\prime\prime}(\bQ_i,\omega) \delta(\bq-\bQ_i) $. One can then replace 
Eq.(\ref{t1}) and (\ref{t2}) by the following analytic expressions:
\beqa
\frac{1}{^{63}T_{1\beta}T} &\simeq& \frac{k_B}{8\pi \hbar } {^{63}F_{\beta}}
(\bQ_i)
\frac{\alpha}{\hbar \omega_{SF}}\\
(1/^{63}T_{2G})^2 &\simeq& \frac{.69}{512}\frac{ ^{63}
F_{ab}^{eff}(\bQ_i)^2 \alpha^2
\xi^2}{ \pi \hbar^2 }.
\eeqa

Another important quantity, the anisotropy
ratio of the $\cu$ spin-lattice relaxation rates, which has been measured for
$\lsco$ and $\ybco$ at various doping concentrations, provides a direct
constraint on  the hyperfine
coupling constants,
$A_{\alpha}$ and $B$. For $\xi \agt 2$, this anisotropy ratio, $\ra$ 
can be written as,
\beq
\ra \equiv \frac{T_{1 c}}{T_{1 ab}} \simeq
\frac{^{63}F_{ab}({\bf Q}_i)}{^{63}F_{c}({\bf Q}_i)}  .
\eeq
For the case of $\lsco$, where the peaks are located at $\bQ_i=(\pi/a, 
[\pi \pm \delta] /a), ([\pi\pm
\delta]/a,\pi/a)$, we then have

\beq
\ra \simeq \frac{1}{2} \left[ 1+ \frac{\Big(A_c - 2B(1+\cos \delta )
\Big)^2}{\Big(A_{ab} - 2B(1+ \cos \delta) \Big)^2} \right].
\label{ralsco}
\eeq
For  $\ybco$,  as indicated in the Introduction, on assuming the broad $\pipi$ peak seen in neutron scattering 
experiments\cite{bourges,mook,keimer,tranquada} reflects 
 the presence of four unresolved overlapping incommensurate 
peaks located along 
the zone
diagonal directions,\cite{mook,tranquada} we may write 
\beq
\bQ_i=([\pi \pm \delta] /a, [\pi
\pm \delta] /a), 
\label{eq14}
\eeq
and the  anisotropy ratio becomes,
\beq
\ra \simeq \frac{1}{2} \left[ 1+ \frac{(A_c - 4B\cos \delta
)^2}{(A_{ab} - 4B\cos \delta )^2} \right].
\label{eq16}
\eeq

Numerical calculations of the $\ox$ relaxation rates 
show that these rates can deviate significantly
 from those obtained by approximating the $\chi_{AF}^{\prime\prime }$ by
a $\delta(\bq-\bQ_i)$ function. We therefore calculate the $\ox$ relaxation rates
numerically, using  Eqs.(\ref{t1}) and (\ref{form}).

\section{The Direct and Transferred Hyperfine Constants}
Seven years of NMR experiments on aligned powders and single crystals of the
cuprates have produced a significant number of constraints which must be taken
into account in selecting the hyperfine constants which enter the SMR
Hamiltonian.  Thus experiments which determine the $\cu$ nuclear resonance
frequency in the $AF$ insulators, $YBa_2Cu_3O_6$\cite{tsuda}  and $La_2CuO_4$\cite{yasuoka}, yield similar
results for the product of $(4B-A_{ab})$ and $\mu_{eff}$, the effective
moment of the localized Cu$^{2+}$ spins,\cite{monien2}
\beqa
\mu_{eff}(4B-A_{ab})& = &  79.65\pm0.05~~kOe  \hspace{.6in}           \mbox{(YBa$_2$Cu$_3$O$_6$)}\\
\mu_{eff}(4B-A_{ab}) & = &   78.78~~kOe     \hspace{1.2in}       \mbox{(La$_2$CuO$_4$)}
\eeqa
On using the value, $\mu_{eff} = 0.62 \mu_B$, determined by Manousakis\cite{manousakis}  for the
2D spin $1/2$ Heisenberg antiferromagnet, we then find
\beqa
4B-A_{ab}& = &128.5~~ kOe/\mu_B            \hspace{1.2in}  (\rm{YBa_2Cu_3O_6})
\label{eq18}\\
4B-A_{ab}& = &127~~ kOe/\mu_B          \hspace{1.2in}    (\rm{La_2CuO_4})
\label{eq19}
\eeqa
A second set of constraints comes from $^{63}Cu$ Knight shift experiments.  
To a
high degree of accuracy, in the $\ybco$ system the $\cu$ Knight shift in a
magnetic field along the $c$-axis is temperature independent in both the
normal and superconducting state, and hence reflects only the chemical shift.
The absence of a spin contribution means that for this system,
\begin{equation}
A_{c} + 4B \simeq 0,
\label{eq21}
\end{equation}
 independent of doping level.  A third set of constraints is
obtained from measurements of the anisotropy of the $^{63}Cu$ spin-lattice
relaxation rates; for $YBa_2Cu_3O_7$ one finds $^{63}R = 3.7\pm 0.1$.\cite{barrett}  To the extent
that $A_{ab}$, $A_{c}$, and $B$ are independent of doping level in $\ybco$,
and the spin fluctuation peaks are commensurate (or nearly so) for this
system, one then finds from Eqs. (\ref{eq16}), (\ref{eq18}), and (\ref{eq21}), that
\beqa
B &=&40.8~~ kOe/\mu_B \nonumber \\
A_{c} &=& -163~~ kOe/\mu_B \nonumber \\
A_{ab} &=&34~~ kOe/\mu_B
\label{ab}
\eeqa
in agreement with the analysis of Monien, Pines, and Takigawa.\cite{monien}
  These values
are consistent with the constraint on $(4B+A_{ab})$ 
obtained by Ishida {\it
et al} for $La_{1.85}Sr_{0.15}CuO_4$; from the slope of a plot of their
direct measurement of
$\chi_o(T)$ against their measured value of $^{63}K_{ab}(T)$, they 
found\cite{ishida}
\begin{equation}
4B+A_{ab} = 189~~ kOe/\mu_B.
\label{eq24}
\end{equation}
It seemed natural therefore to conclude that not only were $A_{ab}$, $A_{c}$,
and $B$ independent of doping for the $\ybco$ system, but that the corresponding
values for the $\lsco$ system were likewise doping independent and were virtually
identical with those deduced for $\ybco$.

If, however, the spin fluctuation peaks in the $\lsco$ system are incommensurate,
the assumption that the hyperfine constraints for this system are doping
independent is no longer tenable for this system, as may be seen by comparing
the measured values of $^{63}R$ for the $\lsco$ system shown in Table I with the
values calculated using Eqs(\ref{ab}), and using the doping dependence of the
degree of incommensuration determined in neutron scattering experiments\cite{aeppli2},
$\delta \sim 1.75 x$, where $x$ is the $Sr$ doping level.  As may be seen in Table
I, the calculated trend with doping is opposite to that seen experimentally.
Since the quantum chemical environment responsible for the direct hyperfine
interaction $A_\alpha$ is not expected to vary substantially with doping, the most likely
culprit in Eqs.(\ref{ab}) is the assumption that the transferred hyperfine coupling
constant does not vary appreciably with doping; indeed, if $B$ increases
sufficiently rapidly with doping, with $A_{ab}$ and $A_{c}$ fixed, one can
find a doping dependence of $^{63}R$ which is more nearly in accord with
experiment.  This means abandoning for the $\lsco$ system the constraint, $A_{c}
\simeq -4B$, which works so well for the $\ybco$ system.

Suppose then one starts anew with the insulator, $La_2CuO_4$.  On making use of
Eqs. (\ref{ralsco}) and (\ref{eq24}) and taking $\ra= 3.9$,  in
accord with the result of Imai $et~al$\cite{imai} at $475$K, one finds readily that
\begin{equation}
A_{ab}-A_{c} = 203~~ kOe/\mu_B.
\label{eq29}
\end{equation}
On turning next to $La_{1.85}Sr_{0.15}CuO_4$, taking $\ra = 3.0$, in
accord with the recent measurement of Milling and Slichter,\cite{milling}
using the result of Ishida {\it et al},\cite{ishida} Eq.(\ref{eq24}), and assuming that
$A_{\alpha\beta}$ is independent of doping, one then finds $B = 48~ kOe/\mu_B$
and $A_{ab} = -3~ kOe/\mu_B$.  This result is, however, unrealistic.  A
straightforward calculation using the expressions adapted by Monien {\it et
al}\cite{monien2} from the work of Bleaney {\it et al}\cite{bleaney},
\beqa
A_{c} &=& 395  [{-\hat{\kappa} - {4 \over 7} - {62 \over 7} \gamma}]~~kOe/\mu_B  \nonumber \\
A_{ab}&=& 395 [{-\hat{\kappa} - {2 \over 7} - {11 \over 7} \gamma}]~~kOe/\mu_B
\label{eq25}
\eeqa
In Eqs.(\ref{eq25}), $\gamma\equiv\lambda/E_{xy}$ is the dimensionless ratio of the spin-orbit coupling for a Cu$^{2+}$ ion, 
$\lambda \sim -710 cm^{-1}$,  to the excitation energy
from the
ground state of the $\cu$ $d_{x^2-y^2}$ orbital of the
various $\cu$ $d$ states, $E_{xy} \sim E_{xz} \sim E_{yz} \sim 2eV$;  with these typical values,
$\gamma = 0.044 \pm 0.009$; $\langle{1 \over r^3}\rangle$ which enters as a
multiplicative factor in Eq. (\ref{eq25}) is taken to be $6.3 a^{-3}_o$.
With the value of $\gamma = 0.0469$ obtained using Eq. (\ref{eq29}),
\begin{equation}
A_{ab} = (-395 \hat{\kappa} + 142)~~ kOe/\mu_B.
\end{equation}
On taking the core polarization $\hat{\kappa} = 0.26 \pm 0.06$\cite{monien2}, we then get, for $\hat{\kappa}$ in the vicinity of its plausible upper limit, 0.32,
\begin{equation}
A_{ab} \geq 16~~ kOe/\mu_B.
\end{equation}

In order to satisfy the above constraints, we next assume that the
anisotropy, $\ra$, for $\lscof$ is at the upper end of the range
quoted by Milling and Slichter, and take $\ra = 3.2$; we next take $A_{ab} = 18~
kOe/\mu_B$ (corresponding to $\hat{\kappa} = 0.316$), a value close, but not at, the
estimated minimum value for $A_{ab}$.  We then have, from Eq. (\ref{eq29}), $A_{c} =
-185~ kOe/\mu_B$ and, from Eq.(\ref{ralsco}) for
$\ra$,
$B_{0.15} = 51~ kOe/\mu_B$, while for the insulator, we find from Eq. (\ref{eq19}), $B_o =
36.1~ kOe/\mu_B$.  With these hyperfine constants we find for
$La_{1.85}Sr_{0.15}CuO_4$ that $4B+A_{ab} = 222 ~kOe/\mu_B$, some $17\%$
above the value obtained by Ishida {\it et al}\cite{ishida} while for 
this system, the ratio of the
spin contributions to the Knight shift for fields parallel and perpendicular
to the c-axis is
\begin{equation}
{^{63}K_{c} \over ^{63}K_{ab}} = {4B+A_{c} \over 4B+A_{ab}} = 8.6\%
\end{equation}
The slight temperature variation of $^{63}K_{c}$ which follows from this
choice of parameters would not be detectable, consistent with the measurements
of Ohsugi $et~al$.\cite{ohsugi}

For intermediate levels of $Sr$ doping, if we assume that the change in $B$
induced by doping scales with the doping level, we obtain the results for
$La_{1.9}Sr_{0.1}CuO_4$ and $La_{1.86}Sr_{0.14}CuO_4$ given in
Table III.  Also given there are the corresponding results for $^{63}T_1$ and
$^{63}T_{2G}$ and related quantities of interest in analyzing NMR
experiments.  We note that to obtain $^{63}R = 3.5$ for
$La_{1.9}Sr_{0.1}CuO_4$, one needs a transferred hyperfine
coupling, $B = 37.8~ kOe/\mu_B$, which is considerably lower than that
obtained by
direct interpolation.

We turn next to the $\ybco$ system.  For $YBa_2Cu_3O_6$, the only constraint on
the hyperfine constants is the AF resonance result, Eq. (\ref{eq18}).  However, as noted
above, for $YBa_2Cu_3O_7$ one has two further constraints:  $4B =
A_{c}$, and $^{63}R = 3.7 \pm 0.1$.\cite{barrett}  Moreover, 
as is the case for
$YBa_2Cu_3O_{6.63}$, neutron scattering experiments on $YBa_2Cu_3O_7$ suggest
that one has four incommensurate and largely unresolved peaks along the zone
diagonal direction whose positions, $\bQ_i$, are given by Eq.(\ref{eq14}).  On
taking $\delta = 0.1$, a value consistent with the experimental results of Dai $et~al$,\cite{mook} we
then find, on making use of Eq. (\ref{eq16}), that
\begin{equation}
A_{ab} = 0.721 B.
\end{equation}
If now we assume that the spin orbit coupling of a $Cu^{2+}$ ion in
$YBa_2Cu_3O_7$ is little changed from that found for $La_2CuO_4$, $\gamma =
0.471$, we have a third relation between the coupling constants,
\begin{equation}
A_{ab}-A_{c} = 4.721 B = 203~~ kOe/\mu_B
\label{eq32}
\end{equation}
from which we find
\beqa
B &=& 43~~ kOe/\mu_B \nonumber \\
A_{ab} &=& 31~~ kOe/\mu_B \nonumber \\
A_{c} &=& -172~~ kOe/\mu_B
\eeqa
while from the $AF$ resonance constraint, Eq. (\ref{eq18}), we find for the insulator
$YBa_2Cu_3O_6$, that $B = 39.8~ kOe/\mu_B$.

Confirmation of this choice of parameters comes by determining the slope from
the linear temperature dependence found in a plot of $^{63}K_{ab}$ versus
$\chi_o(T)$ for $O_{6.63}$.  We find $4B+A_{ab} \sim 200~
kOe/\mu_B$, in  agreement with Eq. (\ref{eq32}).  Moreover, Shimizu {\it et
al}\cite{shimizu} find, from a similar plot for $YBa_2Cu_3O_{6.48}$, that for this
system, $4B+A_{ab} \simeq 200~ kOe/\mu_B$.

We adopt these values in our subsequent calculations.  We note that the value
of $B$ we obtain for $YBa_2Cu_3O_6$ is some $10\%$ larger than that found for
$LaCuO_4$, while the doping dependence of $B$ is considerably smaller in the
$\ybco$ system than in the $\lsco$ system.  Both effects may plausibly
be attributed to the presence of chains in the $\ybco$ system.  The core
polarization parameter, $\hat{\kappa} = 0.281$ we find for the $\ybco$ system is some
$10\%$ smaller than that inferred for the $\lsco$ system.  We tabulate in Table
IV our results for the $\ybco$ system at three doping levels; we estimate $B =
40.6~kOe/\mu_B$ for $YBa_2Cu_3O_{6.63}$ by interpolating between an assumed
value, $B = 39.8$ for $YBa_2Cu_3O_{6.5}$, and that we found above for
$YBa_2Cu_3O_7$.

\section{Reconciling Neutron Scattering and $\cu $ NMR Measurements in
$\lsco$}
We now explore whether, with the revised hyperfine constants proposed above, we
can reconcile the recent neutron scattering results of Aeppli {\it et al}
\cite{aeppli} for $La_{1.86}Sr_{0.14}CuO_4$ with the NMR measurements of 
Ohsugi
{\it et al}\cite{ohsugi} on the two adjacent systems, $La_{1.87}Sr_{0.13}CuO_4$, and
$La_{1.85}Sr_{0.15}CuO_4$.  We assume that $\chi(q,\omega)$ takes the MMP form,
Eq. (\ref{mmp}), in which case
\begin{equation}
\chi''(q,\omega) = \sum_i {\chi_{Q_i} (\omega/\omega_{SF}) \over
[1+(\bQ_i-\bq)^2\xi^2]^2+(\omega/\omega_{SF})^2}
\label{imchi}
\end{equation}
where $\chi_{Q_i}$ is given by Eq.(\ref{eq10}).
There are three undetermined parameters; $\alpha$, $\xi$, and
$\omega_{SF}$.  We begin by deducing $\chi_{Q_i}$ and $\omega_{SF}$ from the
results of Aeppli {\it et al} for $\chi''(Q_i,\omega)$ at $35 K$; as may be
seen in Fig.2, a good fit to their results is found with $\chi_{Q_i}(35K) =
350$ states/eV and $\omega_{SF} = 8.75$meV.  To determine $\alpha$, and hence
$\xi(35K)$, we turn to the NMR results of Ohsugi {\it et al};\cite{ohsugi} on interpolating
between their results for the adjacent systems, as shown in Fig.3, we find
$^{63}T_1T = 34(10^{-3}sK)$, while according to Table III, one has
\begin{equation}
^{63}T_1T = 94.2 (\omega_{SF}/\alpha) sK/(eV)^2.
\label{t1sf}
\end{equation}
Equating these results, we obtain $\alpha = 23.9 states/eV$ and $\xi = 7.6$.

A first check then on our use of Eq. (\ref{imchi}) to fit both NMR and neutron scattering
results is to compare this value of $\xi$ with the measurements of the intrinsic
line width of each peak by Aeppli {\it et al}.\cite{aeppli}  We find on
converting units, that at $35K$ the linewidth parameter of Aeppli {\it et al}
corresponds to a correlation length, $\xi = 7.7$ in the low ($\omega=0$meV) frequency
limit.  The agreement is quite good.

Having determined $\alpha$, we can then use our interpolated NMR results to
obtain $\omega_{SF}(T)$ for $35K \leq T \leq 300K$ from Eq. (\ref{t1sf}).  That leaves
only one parameter, $\chi_{Q_i}$ (or $\xi$) to be determined over this
temperature range.  As a first step toward its determination, we use the results
of Aeppli {\it et al} for $\chi''(Q_i,\omega)$ at $80K$.  As shown in Fig.2, a
good fit to the experimental data is obtained with $\chi_{Q_i}(80K) =
175$ states/eV.  From Eq. (\ref{eq10}), we then get $\xi(80K) = 5.41$.

We next make use of the Barzykin-Pines magnetic phase diagram.  From their
analysis of NMR, transport and static susceptibility experiments, they conclude
that the $\lsco$ system will, like its $\ybco$ counterpart, exhibit non-universal
scaling behavior, perhaps best described as pseudoscaling, between two
cross-over temperatures, $T^*$ and $T_{cr}$.  In this regime, the system
exhibits apparent $z=1$ dynamic scaling behavior, with $\omega_{SF}$ varying
linearly with temperature and
\begin{equation}
\omega_{SF} = c'/\xi
\label{eq40}
\end{equation}
where $c'$ depends on the doping level.  They propose that the upper cross-over
temperature, $T_{cr}$, which marks the onset of pseudoscaling behavior, can be
identified as the maximum in the measured value of $\chi_o(T)$, and corresponds
to a magnetic correlation length, $\xi \sim 2$.  The lower temperature $T^*$ is
 determined from $^{63}T_1$ measurements as the lower limit of the linear
variation of $\omega_{SF}$ (or $^{63}T_1T$) with temperature.  Inspection of
Fig. 3 shows that for $La_{1.86}Sr_{0.14}CuO_4$, one has a comparatively weak
cross-over at $T^* \sim 80K$.  Since $\omega_{SF}(T)$ has already been
determined, a knowledge of $c'$, obtained at one temperature between $T^*$ and
$T_{cr}$, enables one to fix $\xi(T)$ over the entire temperature range.  From
our fit to the neutron data at $80K$, we find $c' = 52.9$meV, and use this
result to conclude that $T_{cr} \sim 325K$, and that
\begin{equation}
{1 \over \xi} = 0.0828 + 0.128 ({T \over 100})   \hspace{1in}  80K < T < 325K
\label{eq41}
\end{equation}
We can interpolate between this result for $\xi(T)$ and our result at $35$K to
obtain $\xi(T)$ over the region, $35K \leq T \leq 300K$.  The result of that
interpolation, which is very nearly a continuation of the linear behavior found
above $80K$, is given in  the inset of Fig. 4.

A first check on the correctness of this procedure is to compare our ``NMR''
derived results at $300K$, shown in Table II, with the neutron scattering
results at this temperature.  As may be seen in Fig.2, the slope, obtained
from the NMR results, $[\chi''(Q,\omega)/\omega] = \chi_{Q_i}/\omega_{SF} =
1.10~\mu_B^2/(eV \cdot meV)$ is in good agreement with experiment.  A second check is to compare our
results for $\xi(T)$ with the values deduced from half-width of the incommensurate peaks in Im$\chi(\bq,\omega)$
observed in neutron scattering over the entire temperature domain $(35 \leq T
\leq 300K)$; that comparison is given in the main portion of Fig.4.   Finally, we can compare the
predictions of Eqs. (\ref{eq40}) and (\ref{eq41}) (the parameters being specified in Table II)
with the combined frequency and temperature dependence of the half-width found
by Aeppli {\it et al} in Fig.5.  In obtaining this 
figure, we calculated the theoretical inverse correlation length $\kappa$ 
from the $\xi(T)$ shown in Fig.4,  by matching the
full width at half maximum of the incommensurate peaks of Eq.(\ref{imchi}) to
 those of the experiments of Ref.\cite{aeppli}. Our comparison of
 the calculated
$\kappa(\omega,T)$ to the experimental values is shown in Fig.5. 
  The extent of the
agreement between our calculations and experiment suggests that we have
succeeded in reconciling the $^{63}Cu$ NMR results with the neutron scattering
results,  and it suggests as well that the neutron scattering results are consistent
with $z=1$ pseudoscaling behavior for temperatures less than $300$K. The latter
conclusion was also reached by Aeppli $et~al$ from their analysis of 
 their neutron scattering experiments.

\section{$\ox$ Relaxation Rates for $\lscof$}
We now demonstrate that by choosing a reasonable next nearest
neighbor hyperfine coupling contribution $C'$, we can reconcile the incommensurate peaks in
$\chi^{\prime\prime}(\bq,\omega)$ with the
measured NMR relaxation rates  $(^{17}T_1T)^{-1}$ for
$\lscof$.



 In calculating  $(^{17}T_{1c}T)^{-1}$ for $\lscof$, 
we simply use
the previously determined parameters as inputs to
Eqs.(\ref{form}) and (\ref{mmp}), where the
 next nearest neighbor Cu-oxygen hyperfine coupling $C^\prime$ is included
in the form
factor $^{17}F_{c}$. For $\lsco$ materials, there is still not enough
experimental data to determine the exact values of $C_{\alpha}^{\rm old}$ for
different field orientations; we therefore assume that
these values are the same as those of the $\ybco$ family.  Following
Monien $et~al$\cite{monien} and Yoshinari $et~al$\cite{yoshinari}, we take
$C_c^{\rm old}=33~kOe$ and $\zeta_\parallel=1.42, ~\zeta_c=1$ and $
\zeta_\perp=0.91$.
We further assume an isotropic $C'$, 
with $r=C'/C_c=0.25$,
and obtain $\chi_0(T)$ by modifying the results of Ref.\cite{barzykin} to reflect the new values of $A_{ab}$ and $B$ presented in Sec.III. We use 
the 
$\alpha$ for $\lscos$ obtained from the neutron scattering fits from 
the last section, and obtain
$\omega_{SF}$, and $\xi(T)$ from NMR data of Ohsugi $et~al$.\cite{ohsugi}
These numbers are almost the the same as those of $\lscos$.  
The remaining parameter
in Eq.(\ref{mmp}),  $\Gamma$, is chosen to get the
best fit to the experimental results for 
 $(^{17}T_{1c}T)^{-1}$. It is important to point out that our  choice
of $\Gamma$  does not affect
 $(^{63}T_{1c}T)^{-1}$ and $1/^{63}T_{2G}$, because the Fermi
Liquid contribution to these quantities is
negligible compared to that of the anomalous antiferromagnetic spin
fluctuations.

In Fig.6 we compare our  calculated  $\ox$  NMR relaxation rate
$(^{17}T_{1c}T)^{-1}$,
using
$\Gamma=345$meV,  with the experimental data of Walstedt $et~al$\cite{walstedt}.
The agreement is quite good. Note, however, the choice of  $r$ and $\Gamma$ is not unique in our
calculations; fits of the same quality can  be obtained by choosing other values
for $r$ and $\Gamma$.
The inset of Fig.6 shows the substantial leakage of the anomalous spin
fluctuations (the first term only in Eq(\ref{mmp}))  to
$(^{17}T_{1c}T)^{-1}$, calculated with the
standard SMR form factor ($r=0$). The
$(^{17}T_{1c}T)^{-1}$ thus calculated  has a 
temperature dependence similar to that of the $\cu$ relaxation rates, 
much faster than seen experimentally. Also shown in the inset is the 
substantially smaller AF leakage calculated from the present hyperfine 
coupling $^{17}F_c$ with $r=0.25$.

\section{Neutron Scattering Line Widths and $\ox$ Spin-Lattice Relaxation
Rates in $\ybco$}

We turn  now to the  neutron scattering and NMR
experiments for the $\ybco$
system. As noted in the introduction,
one apparent problem here has been that the
large $\bq$-width of the antiferromagnetic peak, as observed in the
neutron scattering experiments \cite{bourges,mook,keimer,tranquada},
appeared to be in contradiction with size of the correlation length 
($\xi \agt 2$) required to explain the $\ox$ NMR experiments. As Thelen
and Pines demonstrated \cite{Thelen:Pines}, the half-width 
at half maximum for 
the antiferromagnetic peak in $\chi''({\bf q}, \omega)$
should have been $q_{1/2} \lesssim 0.4/a$ in
order to be consistent with the Mila-Rice-Shastry model and
the oxygen relaxation data for YBa$_2$Cu$_3$O$_7$.
They  found that in order to be consistent with
experiment the leakage from the antiferromagnetic peak
should account for no more than 1/3 of the total measured oxygen
rate. 
This upper bound from NMR is much smaller than
the actual $\bq$-width of the antiferromagnetic peak,
$q_{1/2} \simeq 0.7/a$, observed
in the neutron scattering experiments \cite{mook}.
Assuming the measured width is produced by incommensuration, we plot the
antiferromagnetic ``leakage'' contribution (i.e., that
from the antiferromagnetic part of Eq. (\ref{mmp})) to
the $\ox$ relaxation rate in Fig.7, using the incommensuration
$\delta=0.1 \pi$, which provides a fit to the neutron scattering experiments
\cite{mook}.
Obviously, as in the $\lsco$ material, the temperature
dependence of the measured NMR relaxation rate is remarkably different,
and the amplitude of the ``leakage'' term is too large.
This problem can be avoided by introducing $C'$, as  we have done on
the $\lsco$ system. 
In fact, the much smaller degree of presumed incommensurability
in the $\ybco$ system than that measured directly  for the $\lsco$ system
makes it almost evident that any problem produced by AF leakage can be reconciled by the
same method as used above. We show, in Fig.7, that the AF leakage
contribution for $r = C'/C_c = 0.25$ indeed becomes negligible.
If we assume the same ratio of the AF part to the total rate 
as Thelen and Pines \cite{Thelen:Pines} did, we   
obtain a constraint on $C'$.
We note that the
oxygen form factors Eq.(\ref{nff}) are quadratic in
$\delta {\bf q} = ({\bf q - Q})$
in the vicinity of the antiferromagnetic wave vector ${\bf Q}$:
\beq
^{17}F_{\alpha} = (\delta q_x)^2 \frac{(C_c^{old})^2}{2 (1 + 2 r_c)^2}\,
\sum_{\alpha_i = \alpha', \alpha''} \left[ (2 r_c + 1) \zeta_{\alpha_i} - 4 r_{\alpha_i}
\right]^2 = \eta (\delta q_x)^2
\label{tpp}
\eeq
As a result, the antiferromagnetic contribution to the oxygen 
relaxation rate (which we keep as a constant when we change the form factor) is  
\beq
\left(\frac{1}{^{17}T_1 T}\right)_{AF} \propto \eta ({\bf Q_i - Q})^2,
\eeq
and a change of the oxygen form factor, which alters $\eta$,
produces a new constraint on the acceptable width (or incommensurability)
of the neutron scattering peak.
Since Thelen and Pines \cite{Thelen:Pines} used the isotropic form
of the Mila-Rice-Shastry Hamiltonian, with $C_{iso}=C_c^{old}$, we easily
obtain from Eq.(\ref{tpp}):
\beq
q_{1/2 \alpha} \lesssim 0.4/a \frac{1 + 2 r_c}{\sqrt{\frac{1}{2}\,\sum_{\alpha_i =
 \alpha', \alpha''} \left[ (2 r_c + 1) \zeta_{\alpha_i} - 4 r_{\alpha_i}
\right]^2}},
\label{tpp1}
\eeq
where we have neglected possible slow (logarithmic) dependence. In particular,
with  $r_{\alpha_i}=0.25$, Eq.(\ref{tpp1}) gives the upper limit:
$q_{1/2 } \lesssim 0.7/a$. This crude estimate shows that indeed, our
hyperfine Hamiltonian is consistent with both NMR and neutron scattering
experiments. However, the antiferromagnetic leakage contribution to the oxygen
relaxation rate in $\ybco$ can become important,
and should therefore be calculated numerically, since the spin-spin correlation
length is very short.

  For our numerical calculation of the antiferromagnetic peak contribution
to the $\ox$ relaxation rates we assume, as indicated in the Introduction, 
that the neutron scattering data of Tranquada {\em et al}
\cite{tranquada} and Dai {\em et al} \cite{mook} can be interpreted as
indicating that the magnetic response
function $\chi({\bf q}, \omega)$ possesses four incommensurate peaks located
at ${\bf Q}_i = (\pi \pm \delta, \pi \pm \delta)$, and take 
$\delta \simeq 0.1 \pi$, an incommensuration consistent with the measured experimental widths. 
We also assume that the
temperature-dependent spectral weight for these incommensurate peaks, as
in case of $\lsco$, comes from the temperature dependence of the correlation
length, and adopt the MMP form Eq.(\ref{mmp}) for each of the four
peaks. It should be emphasized, however, that accord between the
inelastic neutron scattering and the oxygen NMR can be reached for any
bell-shaped curve for $\chi''({\bf q},\omega)$ which has the characteristic
width  measured in the neutron scattering experiments, and a
sufficiently abrupt fall-off at large $({\bf q} - {\bf Q})$.
In Fig.7, we show our calculated antiferromagnetic
leakage to the oxygen relaxation $^{17}W_{1c}/T\equiv 1.5 (^{17}T_{1c}T)^{-1}$, for the case of both $r=0$ and
$r=0.25$; again, we see that the new form factor with $r=0.25$ greatly
reduces the AF leakage. Also shown in Fig.7 is our calculated $^{17}W_{1c}/T$
plotted against the experimental data of Martindale $et~al$.\cite{martindale}
In obtaining our theoretical result, we have used as an input to Eq.(\ref{mmp}), 
 $\chi_0(T)$ deduced from 
the Knight shift $K_c(T)$ data on the same sample, provided by Martindale
$et~al$\cite{martindale2}  and used the $\xi$ and
$\alpha$ from Ref.\cite{barzykin}. Again, we take $C_c^{\rm old}=33~kOe/\mu_B$
for $\ybco$ system.
 By assuming $r_{\alpha_i}=0.25$, we obtain a good fit
to the experimental data with $\Gamma\sim 308meV$. These parameters are listed 
in Table IV.

  Another problem with the one-component Shastry-Mila-Rice picture
has been pointed out recently by Martindale {\em et al}\cite{martindale}, who measured planar $\ox$
relaxation rates for different magnetic field directions.
They have found that the {\em temperature dependences} of the relaxation
rates for magnetic fields parallel and perpendicular to the Cu-O bond axis directions were different, in
contradiction with the predictions based on the MSR hyperfine Hamiltonian for which 
the oxygen form factor is given by
Eq.(\ref{form}), without $C'$:
\beq
^{17}F_\alpha = 2\sum_{\alpha_i= \alpha^{\prime}, \alpha^{\prime\prime}}
  \cos^2 \frac{q_xa}{2}C_{\alpha_i}^2.
\label{oldf}
\eeq
  From Eq.(\ref{oldf}) it follows that the ratios of the oxygen relaxation rates for different
magnetic field orientations should be temperature-independent, and
determined only by the Hyperfine $C$-couplings:
\beq
\label{ora}
\frac{^{17}(1/T_{1, \alpha_i})}{^{17}(1/T_{1, \alpha_j})}\, =
\frac{C_{\alpha_i'}^2 + C_{\alpha_i''}^2}{C_{\alpha_j'}^2 + C_{\alpha_j''}^2}\,
\eeq
Experimentally, as shown by Martindale {\em et al},\cite{martindale} these ratios turn out
to be mildly temperature-dependent, although numerically close to the values
of Eq.(\ref{ora}) .

This apparent contradiction can, in fact, be turned
into a proof of the validity of the modified one-component
model Eq.(\ref{newh}).
It can easily be seen that for our oxygen form factors,
Eq.(\ref{nff}), the $\ox$ relaxation rates
for different field directions do not have the same $\bq$-dependence
for the whole Brillouin zone.
As a result,  ratios such as Eq.(\ref{ora}) should indeed become
temperature-dependent. 
Since we do not know the precise values of the  
couplings once we go beyond the nearest-neighbor 
Mila-Rice-Shastry approximation, we use here the 
expressions for the oxygen form-factors in the most general
form. 
To derive the form of the temperature dependence,
we separate the antiferromagnetic and the Fermi liquid or 
short wave-length ($\chi_0$)
contributions to $(1/^{17}T_{1\alpha})$, according to Eq.(\ref{mmp}):
\beq
\frac{1}{^{17}T_{1\alpha} T}\,
= \left(\frac{1}{^{17}T_{1\alpha} T}\,\right)_{\chi_0} +
 \left(\frac{1}{^{17}T_{1\alpha} T}\,\right)_{AF}
\label{ttsep}
\eeq
Here the short wave-length part $(1/^{17}T_1 T)_{\chi_0}$ is proportional to
the bulk magnetic susceptibility $\chi_0(T)$, while the antiferromagnetic
part follows the copper relaxation rate:
\beqa
\nonumber
\left(\frac{1}{^{17}T_{1\alpha} T}\,\right)_{\chi_0} & = 
& S_{\alpha} \chi_0(T) \\
\left(\frac{1}{^{17}T_{1\alpha} T}\,\right)_{AF} & = &
\frac{^{17}F_{\alpha}({\bf Q_i})}{^{63}F_{c}({\bf Q_i})}\,
\left(\frac{1}{^{63}T_{1c} T}\,\right) 
\eeqa
As we have demonstrated above, the temperature-dependence of the
antiferromagnetic leakage term is very different from what is
observed in experiment.
Since the empirical modified Korringa law
$(1/^{17}T_1 T \chi_0(T)) = const$ is rather well satisfied for these
materials, the short wave-length part should be dominant. 
Therefore, we can write for the different $\ox$ relaxation rate ratios:
\beq
\label{oxle}
\frac{(1/^{17}T_1)_{\alpha_i}}{(1/^{17}T_1)_{\alpha_j}}\,  =
\frac{S_i \chi_0(T) + \frac{P_i}{^{63}T_{1c} T}\,}{
S_j \chi_0(T) + \frac{P_j}{^{63}T_{1c} T}\,}\, \simeq
\frac{S_i}{S_j}\,\left[1 + \left(\frac{P_i}{S_i}\,-\frac{P_j}{S_j}\,\right)
\left(\frac{1}{^{63}T_{1c} T \chi_0(T)}\,\right) \right],
\eeq
 where $P_j = {^{17}F_{\alpha_j}({\bf Q_i})}/{^{63}F_{c}({\bf Q_i})}$,
while the $S_j$ are coefficients determined by  integrating  the product of the
short-range part of the magnetic susceptibility with the oxygen form factor.
If the short-range part of $\chi''({\bf q}, \omega)$
is only mildly $\bq$-dependent,  $S_j$ is determined primarily by the momentum 
average of $^{17}F_{\alpha j}$,
\beq
S_j = \frac{\pi}{\Gamma} \, \int \ ^{17} F_{\alpha_j}({\bf q})
d^2 q
\label{kij}
\eeq
 In this case the temperature-independent part of the ratio
of the oxygen relaxation rates  is determined again only by the
ratio of the form factors:
\beq
\frac{S_i}{S_j}\, = \frac{^{17} F_{\alpha_i}({\bf r}=0)}{^{17}
F_{\alpha_j}({\bf r}=0)}
\eeq
If a realistic band-structure $\bq$-dependence of $\chi''({\bf q},\omega)$ 
is taken into account, this ratio will have a somewhat different value.
We demonstrate, in Fig.8, that expression Eq.(\ref{oxle}) indeed provides
a consistent explanation of the temperature-dependent term for the oxygen
relaxation rates in YBa$_2$Cu$_3$O$_7$; on using Eq.(\ref{oxle}) to fit
the observed anisotropy ratio of $^{17}W_{\parallel} /^{17}W_{\perp}$,
we find
\beqa
\frac{S_\parallel}{S_\perp}&=&0.5, \nonumber \\
\frac{S_\parallel}{S_\perp}\left( \frac{P_\parallel}{S_\parallel}
-\frac{P_\perp}{S_\perp} \right)&=&0.06  ~~(sK)  \mu_B^2/eV
\eeqa
These values of $S_{i,j}$ and $P_{ij}$  impose certain constraints
on the parametric space of the hyperfine couplings. 
However, there are not enough of these
constraints to enable us to deduce unambiguously 
the values of the hyperfine couplings, so that
 specific quantum
chemical calculations are needed to determine the hyperfine coupling
constants for these materials. However, as we have shown,
the temperature dependence of the rates can be accounted for by assuming
 a finite incommensurability for the antiferromagnetic peak.

Using our formalism, and the constants $S_j$ and $P_j$ for $\ybcos$,
we can predict the oxygen relaxation rates for $\ybcox$. It is
easy to see that if the hyperfine C-couplings do not depend significantly
on doping, the product, $S_j \Gamma$, for $\ybcox$ is the same as for $\ybcos$. $P_j$,
however, can be somewhat different, corresponding to a different amount
 of incommensuration for $\ybcox$. Since the oxygen form factor
is quadratic in the vicinity of $\pipi$, we can write:
\beq
P_{j\ybcox} = P_{j\ybcos} \frac{\delta^2_{\ybcox}}{\delta^2_{\ybcos}}
\eeq
However, the degree of incommensurability in $\ybcox$ \cite{tranquada}
is roughly the same as in $\ybcos$, both have $\delta\sim 0.1\pi$, 
so that $P_j$ will remain unchanged 
from the $\ybcos$ values.
This makes it possible to predict 
 the behavior of the oxygen relaxation rates
ratios in $\ybcox$, once the parameter $\Gamma$ in Eq.(\ref{mmp}) 
is determined from experiment. In Fig.9, we fit the $\ox$ relaxation rates
$(^{17}T_{1c}T)^{-1}$ of Takigawa $et~al$ to determine $S_c$. 
Again, we use the $\chi_0$, $\alpha$ and $\xi$ given in Ref.\cite{barzykin}. 
It is seen in the main portion of Fig.9 that the fit is very satisfactory, 
from this fit we obtain $\Gamma=226meV$ for $\ybcox$, if we assume $r_{\alpha_i}=0.25$. 
These parameters are
also listed in Table IV. From Eq.(\ref{kij}), 
we have $S_i\Gamma$ being the same in $\ybcos$ and $\ybcox$, because their
form factors do not change. Therefore, we get for $\ybcox$,
\beqa
& &\frac{S_\parallel}{S_\perp}=0.5, \nonumber \\
\frac{S_\parallel}{S_\perp}\left( \frac{P_\parallel}{S_\parallel}
-\frac{P_\perp}{S_\perp} \right) _{\ybcox}&=&
\frac{S_\parallel}{S_\perp}\left( \frac{P_\parallel}{S_\parallel}
-\frac{P_\perp}{S_\perp} \right) _{\ybcos} \times \frac{S_c(\ybcos)}
{S_c(\ybcox)} \nonumber \\
&=&
0.06 \times \frac {\Gamma_{\ybcox}}{\Gamma_{\ybcos}}= 0.044 ~~(sK)  \mu_B^2/eV
\eeqa
We show our calculated relaxation
rate ratio for $\ybcox$ in the inset of Fig.9.

\section{Discussion and Conclusions}
We have seen that by modifying the SMR hyperfine Hamiltonian we can use the
 MMP one-component spin-spin response function to reconcile the results of a 
number of  neutron scattering and NMR experiments on the cuprate 
superconductors.  With the aid of the scaling arguments of Barzykin and 
Pines,\cite{barzykin} we are able to obtain a quantitative fit to both 
the NMR and the neutron scattering data for $\lscos$. We find that for 
the $\ybco$ system, we can reconcile the $\bq$-width of the antiferromagnetic 
peak seen in neutron scattering experiments with the substantial temperature 
dependent AF correlation required to explain the NMR experiments on $\ybcos$ 
and $\ybcox$. Moreover, in the recent results of Martindale $et~al$\cite{martindale} on the
anomalous temperature dependence of the anisotropy of the $\ox$ relaxation 
rates, the  small amount
of the AF leakage is shown not only to be explicable using our modified one-component 
description; but to provide a direct 
proof for the one-component picture.
Our ability to reconcile so many different experiments leads us to conclude 
that a transferred hyperfine coupling between next nearest neighbor Cu$^{2+}$ 
spins and $\ox$ nuclei spin plays a significant role, and that the transferred 
hyperfine coupling B, changes as one goes from the $\lsco$ to the $\ybco$ 
system, and is moreover comparatively sensitive to hole doping in the former 
system. It will be interesting to see whether the presence of these terms 
can be justified microscopically through detailed quantum chemical 
calculations in these systems.

Our results have a number of interesting implications for NAFL (nearly 
antiferromagnetic  Fermi liquid theory) calculations of other properties 
of the superconducting cuprates. For example, Pines and Monthoux\cite{pines} 
have shown that incommensuration acts to lower the superconducting transition 
temperature, $\Tc$; it is tempting therefore to attribute much of the 
substantially difference in $\Tc$ found for the $\lsco$ and $\ybco$ systems 
to the much greater degree of incommensuration found in the former materials. 
In their calculation of planar resistivities, Stojkovic and Pines\cite{branko} 
find that $\rho_{ab}$ depends sensitively on the size and distribution of 
``hot spots''( regions of the Fermi surface connected by $\bQ_i$), and thus 
is markedly changed by incommensuration. To cite a third example, in NAFL 
theory, the location in momentum space of the peak in the spin fluctuation 
spectrum depends on the interplay of the peaks in the 
irreducible particle-hole susceptibility, $\tilde{\chi} (\bq,0)$,
 produced by band structure 
and the momentum-dependence of the restoring force, $J_{\bq}$, which 
acts to shift those peaks according to Ref.\cite{ref38}, 
\beq
\chi(\bq, 0)=\frac{\tilde{\chi}(\bq,0)}{1-J_{\bq} \tilde{\chi}(\bq,0) }.
\eeq
Since the peaks in $ \tilde{\chi}(\bq,0) $ move away from $\pipi$ as one 
moves away from half-filling, less peaking in $J_{\bq}$ is required to 
produce four incommensurate peaks than was needed by Monthoux and 
Pines\cite{ref38} to keep the peak at $\pipi$ in the presence of 
substantial hole doping. 

Further NMR and neutron experiments on the $\ybco$ and $\lsco$ systems can 
also help verify the correctness of our proposed new hyperfine Hamiltonian 
and our assignment of incommensurate peaks in the $\ybco$ system. For 
example, our results, Eq.(\ref{tpp}) and Eq.(\ref{oxle}), lead us to predict 
substantial temperature dependence in the anisotropy of $1/^{17}T_{1\alpha}$ 
in the $\lsco$ system for magnetic fields parallel and perpendicular to the 
Cu-O bond axis, and it will be instructive to see whether this can be measured.
 It is, moreover, to be hoped that improvements both in neutron scattering 
facilities and the availability of large single crystals will make possible 
a direct experimental check on our assignment of incommensuration in the 
$\ybco$ system. Resolution of those peaks, together with a direct measurement 
of their intensities would also enable one to carry out a detailed comparison 
of NMR and neutron scattering experiments on $\ybcox$ analogous to that 
presented here for $\lscos$ system.

\acknowledgments
We would like to thank G. Aeppli, T. Mason, J. Martindale, C. Hammel, 
C. P. Slichter and C. Milling for communicating their results to us in advance 
of publication, and for stimulating discussions on these and related topics. 
This work is supported in part (YZ and DP) by NSF-DMR-91-20000 through the Science and
Technology Center for Superconductivity and in part (VB) by NSERC of Canada. 
Two of us (VB and DP) wish to thank the
Aspen Center for Physics, where part of this work was carried out, 
for its hospitality.


\newpage
\begin{table}
\caption{Spin-lattice anisotropy and incommensuration in the $\lsco$ system}
\begin{tabular}{|c||c|c|c|c|c|}
System&$\delta$&($^{63}R)_{expt}$&Ref.&($^{63}R)_{Eq.(\protect\ref{ab})}$
&$^{63}R_{Eq.(\protect\ref{ralsco} )}$\\ \hline
$LaCu_2O_4$&0&3.9$\pm0.3$& \cite{imai} &3.7&3.9\\
$La_{1.9}Sr_{0.1}Cu_2O_4$&.175&3.5$\pm?$&\cite{ohsugi}&4.11&3.2\\
$La_{1.85}Sr_{0.15}Cu_2O_4$&.263&$3.0\pm0.2$0.&\cite{milling}&4.78&3.2
\end{tabular}
\end{table}

\begin{table}
\caption{Fits to neutron scattering experiments of
${\rm La_{1.86}Sr_{0.14}CuO_4}$}
\begin{tabular}{|c||c|c|c|}
&T=35K&T=80K&T=325K\\ \hline\hline
${\rm ^{63}T_1T (\times 10^{-3}sK)}$ &34&38&103\\
$\omega_{SF}$(meV)&8.75&9.78&24.7\\
$\alpha  {\rm (states/eV)}$&23.9&23.9&23.9\\
$\chi_{\bQ} {\rm (\mu_B^2/eV)}$&350&175&26.2\\
$\xi$ (a)&7.6&5.41&2.14\\
${c'}$&&52.9&\\ \hline
${\rm 1/T_{2G} (msec^{-1})}$&63&45&18
\end{tabular}
\end{table}
\begin{table}
\caption{Parameters for $\lsco$}
\begin{tabular}{|c||c|c|c|c|}
&${\rm La_{2}CuO_4}$&${\rm La_{1.90}Sr_{0.10}CuO_4}$
&${\rm La_{1.86}Sr_{0.14}CuO_4}$&${\rm La_{1.85}Sr_{0.15}CuO_4}$
\\ \hline
${\rm A_c}$(kOe/$\mu_B$)&-185&-185&-185&-185\\
${\rm A_{ab}}$(kOe/$\mu_B$)&      18&18&18&18\\
B(kOe/$\mu_B$)&36.1&46&50&51\\
C$_c$(kOe/$\mu_B$)&33&33&33&33\\
${\rm ^{63}R_{exp}}$&3.9$\pm$0.3&3.5$\pm?$&&3.0$\pm0.2$\\
${\rm ^{63}R_{cal}}$&3.9&3.2&3.2&3.2\\
4B-${\rm A_{ab}}$(kOe/$\mu_B$)&127&166&182&186\\
4B+${\rm A_{ab}}$(kOe/$\mu_B$)&162.4&202&218&222\\
${\rm ^{63}{K_{c}}/{^{63}{K_{ab}}}}$&-25\%&-0.5\%&7\%&8.6\%\\
${\rm ^{63}T_1T}$&138$(sK/eV^2)\omega_{SF}/\alpha$&95.2$(sK/eV^2)\omega_{SF}/\alpha$
&93.5$(sK/eV^2)\omega_{SF}/\alpha$&94.2$(sK/eV^2)\omega_{SF}/\alpha$\\
${\rm {1}/{T_{2G}}}$&298(eV/s)$\alpha\xi$&347(eV/s)$\alpha\xi$&350(eV/s)$\alpha\xi$&348(eV/s)$\alpha\xi$\\
${\rm {^{63}T_1T}/{T_{2G}}}$&4.12$\times 10^4 ({K}/{eV})\omega_{SF} \xi$&
3.30$\times 10^4 ({K}/{eV})\omega_{SF} \xi$&3.27$\times 10^4 ({K}/{eV})\omega_{SF} \xi
$&3.28$\times 10^4 ({K}/{eV})\omega_{SF} \xi$\\
${\rm {^{63}T_1T}/{T_{2G}^2}}$
&1.23$\times 10^7 ({K}/{s})\alpha \omega_{SF}\xi^2$
&1.15$\times 10^7 ({K}/{s})\alpha \omega_{SF}\xi^2$
&1.14$\times 10^7 ({K}/{s})\alpha \omega_{SF}\xi^2$
&1.14$\times 10^7 ({K}/{s})\alpha \omega_{SF}\xi^2$\\ \hline
$\Gamma$ (meV)&&&&345\\
$r$&&&&0.25\\
$\delta$&&0.175&0.245&0.263\\
\end{tabular}
\end{table}

\begin{table}
\caption{Parameters for $\ybco$}
\begin{tabular}{|c||c|c|c|}
&${\rm YBa_2Cu_3O_{6}}$&${\rm YBa_2Cu_3O_{6.63}}$
&${\rm YBa_2Cu_3O_{7}}$ \\ \hline
${\rm A_c}$(kOe/$\mu_B$)&-172&-172&-172\\
${\rm A_{ab}}$(kOe/$\mu_B$)& 31&31&31\\
B(kOe/$\mu_B$)&39.8&40.6&43\\
C$_c$(kOe/$\mu_B$)&33&33&33\\
${\rm ^{63}R_{exp}}$&&&3.7$\pm0.1$\\
${\rm ^{63}R_{cal}}$&3.8&4.0&3.7\\
4B-${\rm A_{ab}}$(kOe/$\mu_B$)&128.5&131.4&141\\
4B+${\rm A_{ab}}$(kOe/$\mu_B$)&190&193&203\\
${\rm ^{63}{K_c}}/{^{63}K_{ab}}$&-7\%&-5\%&0\\
${\rm ^{63}T_1T}$&135$(sK/eV^2)\omega_{SF}/\alpha$
&145$(sK/eV^2)\omega_{SF}/\alpha$&126$(sK/eV^2)\omega_{SF}/\alpha$\\
${\rm {1}/{T_{2G}}}$&301(eV/s)$\alpha\xi$&293(eV/s)$\alpha\xi$
&310(eV/s)$\alpha\xi$\\
${\rm {^{63}T_1T}/{T_{2G}}}$&4.06$\times 10^4 ({K}/{eV})\omega_{SF} \xi$&
4.25$\times 10^4 ({K}/{eV})\omega_{SF} \xi$
&3.9$\times 10^4 ({K}/{eV})\omega_{SF} \xi$\\
${\rm {^{63}T_1T}/{T_{2G}^2}}$
&1.22$\times 10^7 ({K}/{s})\alpha \omega_{SF}\xi^2$
&1.25$\times 10^7 ({K}/{s})\alpha \omega_{SF}\xi^2$
&1.21$\times 10^7 ({K}/{s})\alpha \omega_{SF}\xi^2$ \\ \hline
$\alpha$ &&8.34&15.36\\
$\Gamma$ (meV)&&226&308\\
$r$&&0.25&0.25\\
$\delta$&&0.1&0.1\\
\end{tabular}
\end{table}


 \begin{figure}
\end{figure}

\begin{figure}
\centerline{\epsfxsize=15cm \epsfbox{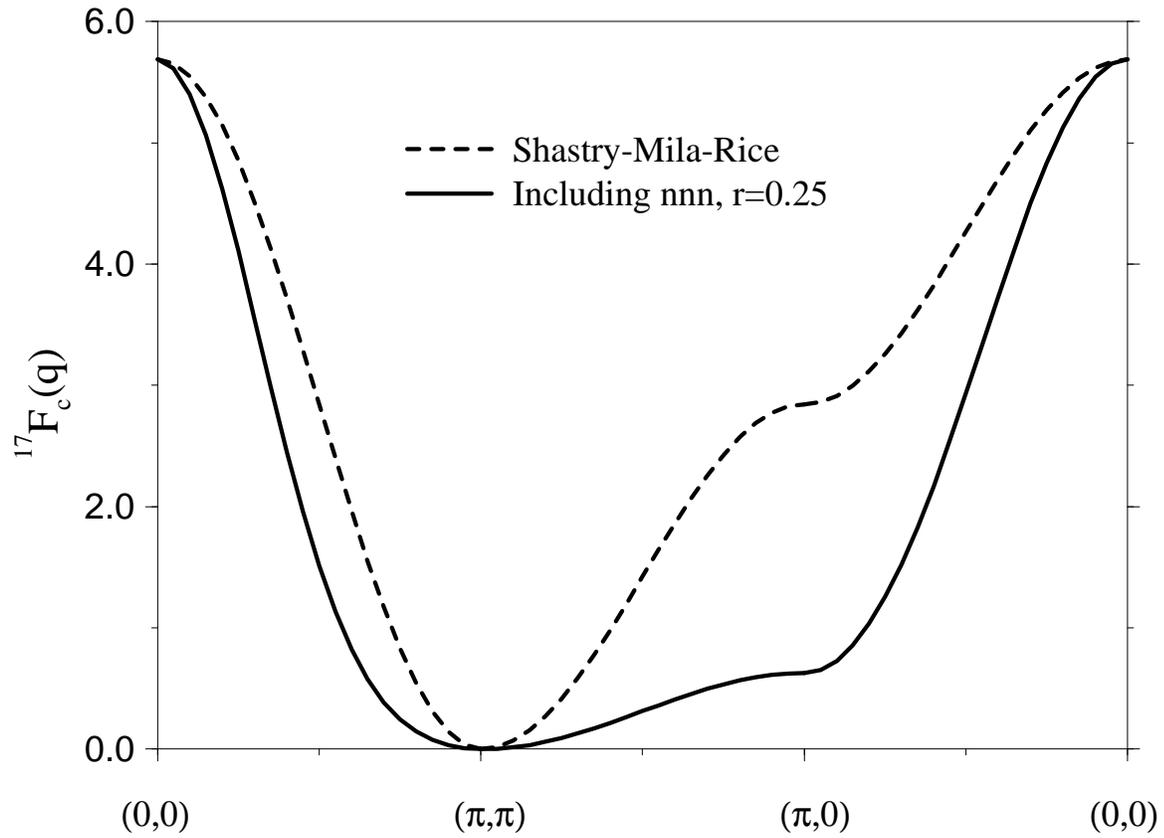}}\vspace{-2in}
\caption{Comparison of the modified form factor  of  $^{17}F_c $
in Eq(\protect\ref{nff}) with $r=0.25$ (solid line) with the standard
Shastry-Mila-Rice
 form (dashed line). $^{17}F_c$ is plotted in units of $(C_c^{\rm old})^2$.
}
\end{figure}

\begin{figure}
\centerline{\epsfxsize=15cm \epsfbox{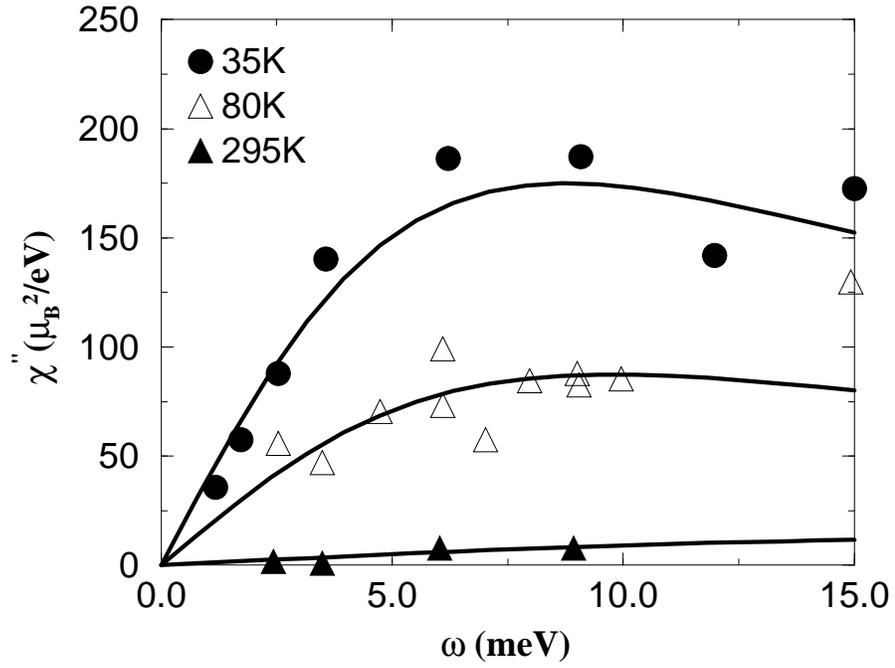}}\vspace{-2in}
\caption{The frequency dependence of $\chi^{\prime\prime}(\bQ_i,\omega)$ at 
three temperatures. The experimental points are the results obtained by Aeppli $et~al$\protect\cite{aeppli}; the solid curves are the fits obtained using a mean
field description of $\chi''(\bq,\omega)$ shown in Eq.(\protect\ref{imchi})
and parameters compatible with NMR results.
}
\end{figure}

\begin{figure}
\centerline{\epsfxsize=15cm \epsfbox{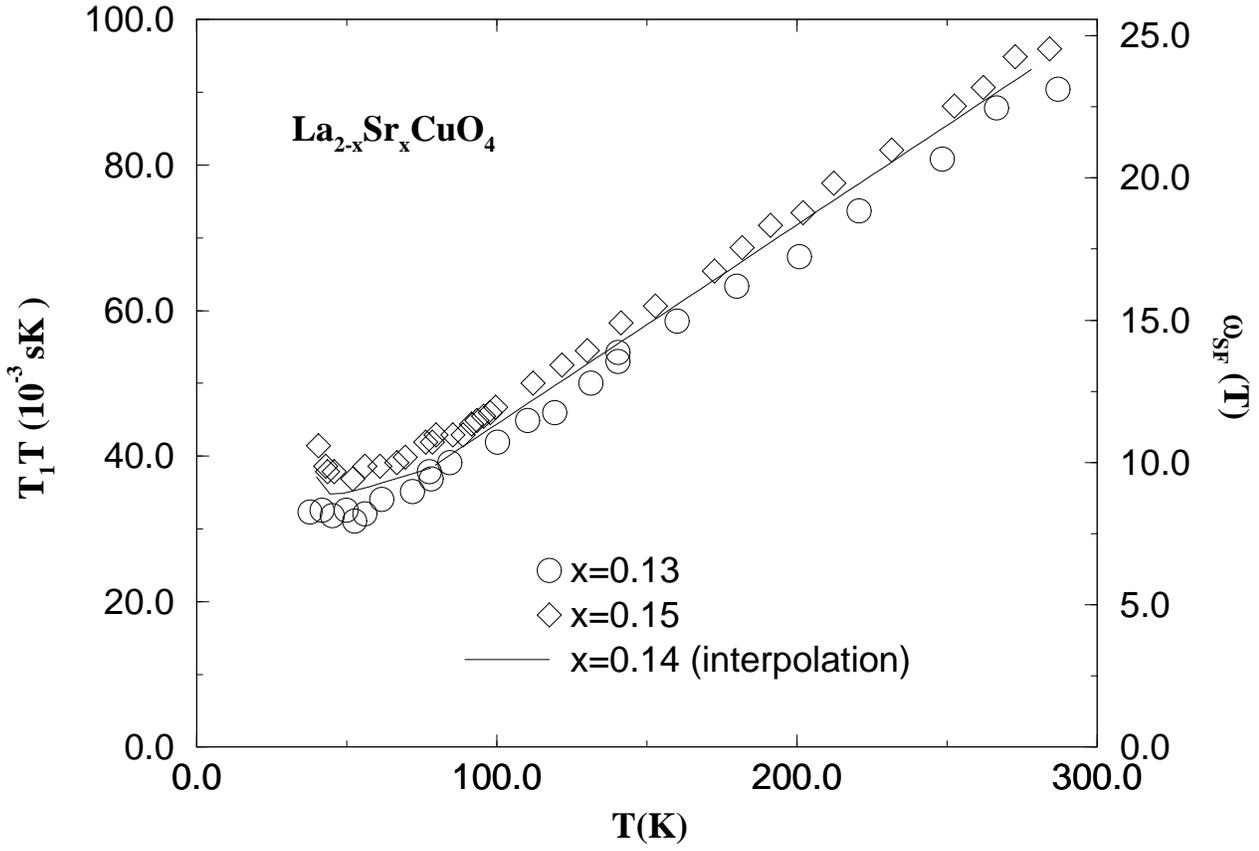}}\vspace{-2in}
\caption{ 
The interpolated $^{63}T_{1c}T$ for ${\rm La_{1.86}Sr_{0.14}CuO_4}$ is shown
together with the measured values of $^{63}T_{1c}T$ for 
${\rm La_{1.87}Sr_{0.13}CuO_4}$ and $\lscof$ of Ohsugi 
$et~al$.\protect\cite{ohsugi} Shown on the right hand side is 
the scale for $\omega_{SF}(T)\propto {^{63}T_{1c}}T$ for $\lscos$ inferred from the 
fit to the neutron scattering experiments.
}
\end{figure}

\begin{figure}
\centerline{\epsfxsize=15cm \epsfbox{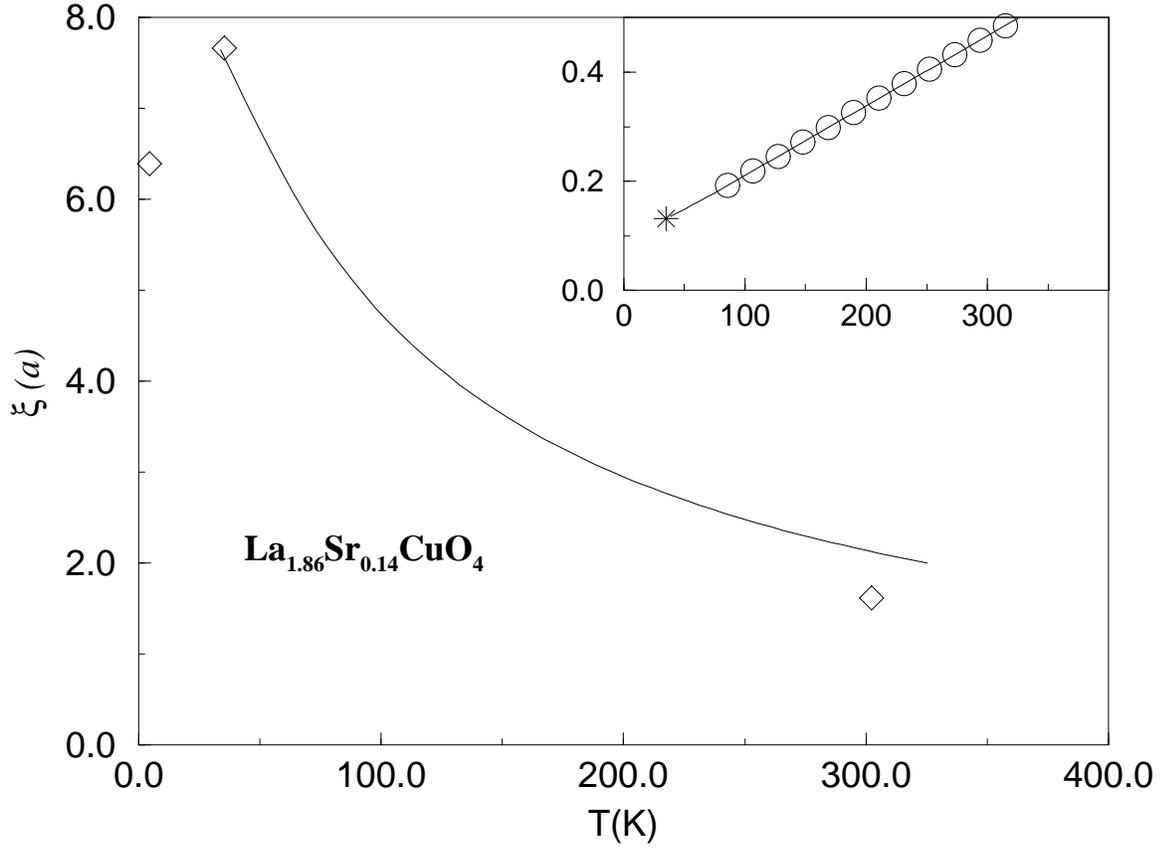}}\vspace{-2in}
\caption{A comparison of the NMR-deduced values of $\xi(T)$ (solid line)
 for $\lscos$ 
with those obtained (diamonds) from the neutron scattering experiments of
Ref.\protect\cite{aeppli}. The experimental points (diamonds) are derived 
by first fitting the half-width of the neutron scattering peak at $\bQ_i$ for low energy transfer  ($\omega=2.5meV$), and then
extrapolating to $\omega=0$, following the formula 
$\kappa^2(\omega,T)=\kappa^2(T)
+ a^{-2}\omega^2/E_\omega^2$ given in Ref.\protect\cite{aeppli}.
 The inset shows the interpolation procedure used to obtain
$1/\xi(T)$ between 35K and 80K: the point at 35K (star) is obtained from the MMP fit
to the neutron scattering data at 35K, while the points above 80K are
 deduced from
the scaling analysis of Eq.(\protect\ref{eq41}) (circles); the solid line
shows the  extrapolation between 35K and 80K.
}
\end{figure}
\begin{figure}
\centerline{\epsfxsize=15cm \epsfbox{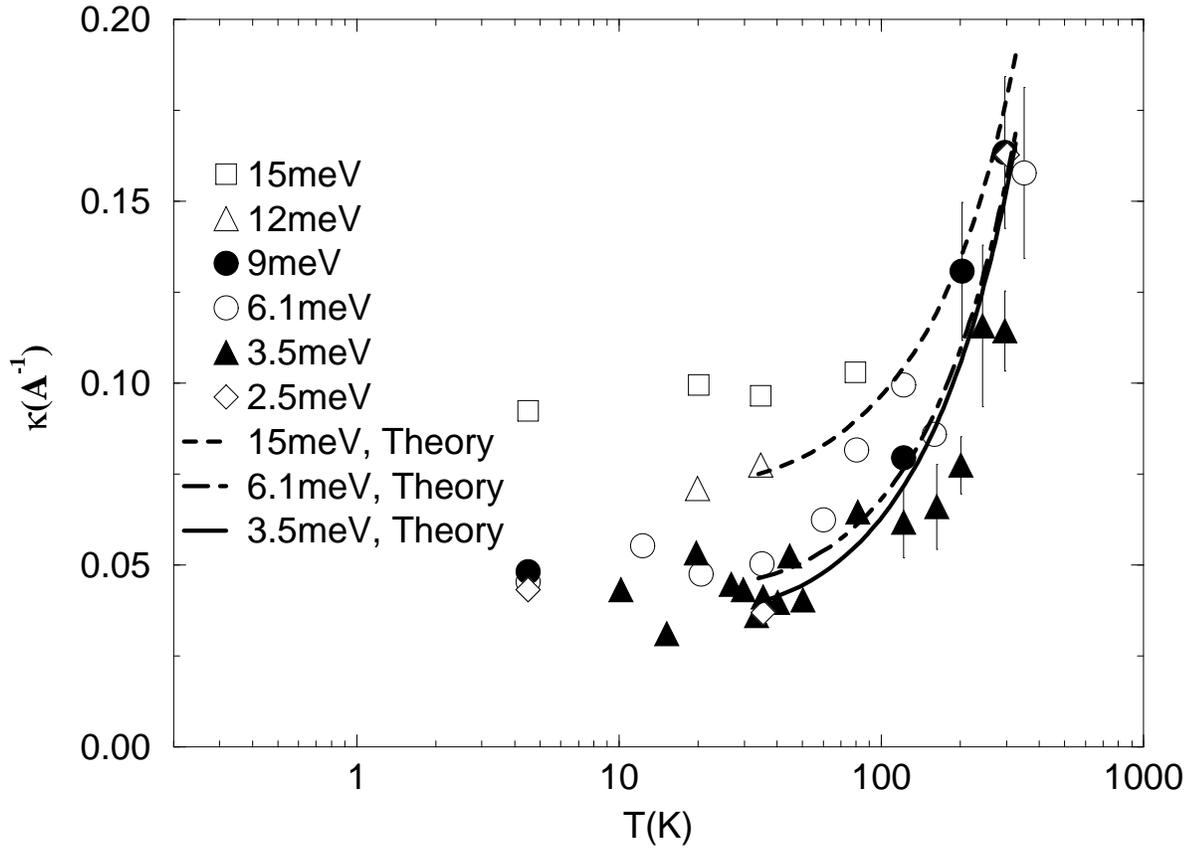}}\vspace{-2in}
\caption{A comparison of the NMR-deduced values of the frequency dependence 
inverse correlations length, $\kappa$, at $\omega$=3.5, 6.1, and 15meV 
 (lines), with the experimental results of Aeppli 
$et~al$\protect\cite{aeppli} (symbols). It is seen that the consistency is 
quite good at low frequencies, while
at high frequencies, the NMR-deduced $\bq$-width is smaller than those
seen in neutron scattering experiments.
}
\end{figure}

\begin{figure}
\centerline{\epsfxsize=15cm \epsfbox{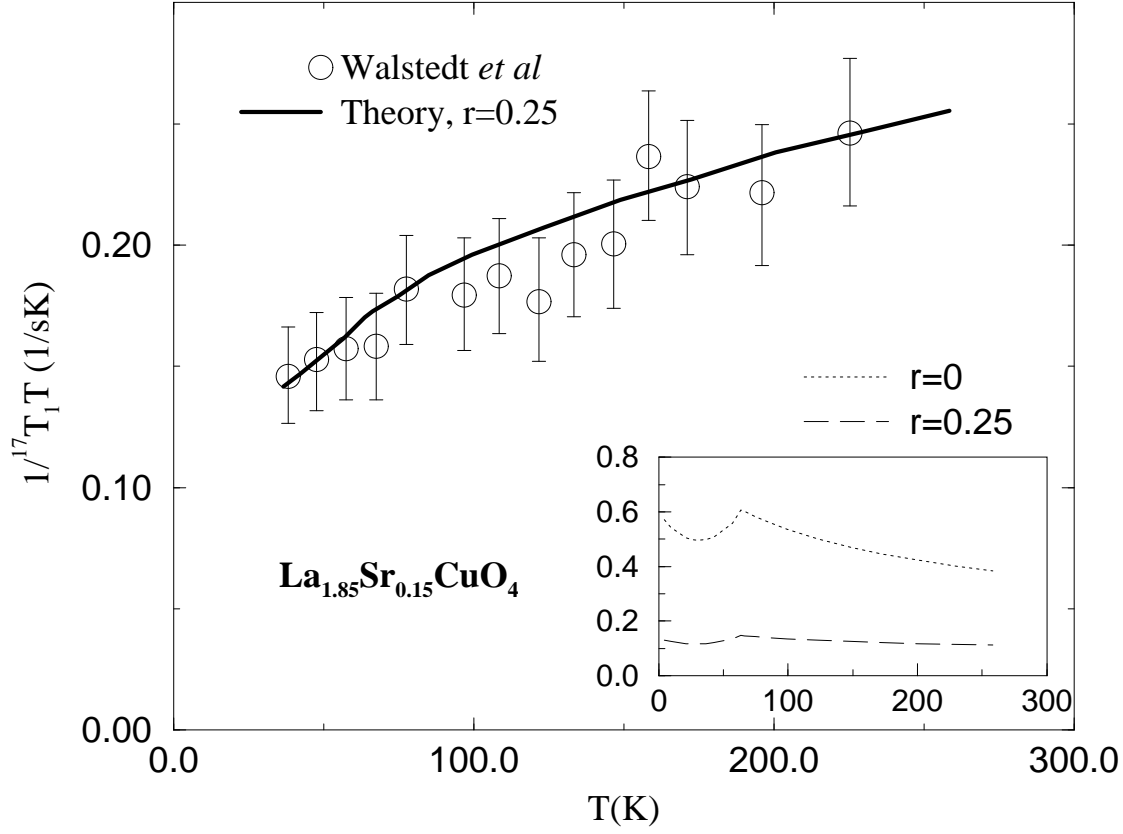}}\vspace{-2in}
\caption{Our calculated spin-lattice relaxation rates $(^{17}T_{1c}T)^{-1}$ 
for $\ox$ (solid line) compared to the experimental data of Walstedt 
$et~al$\protect\cite{walstedt} (circles) for $\lscof$; we use r=0.25 for 
the calculation. The inset shows the 
contribution of the AF spin fluctuations (first term only in 
Eq.(\protect\ref{mmp})) to 
$(^{17}T_{1c}T)^{-1}$, calculated with the present form factor (r=0.25) and
with the standard Shastry-Mila-Rice form factor (r=0).
}
\end{figure}

\begin{figure}
\centerline{\epsfxsize=15cm \epsfbox{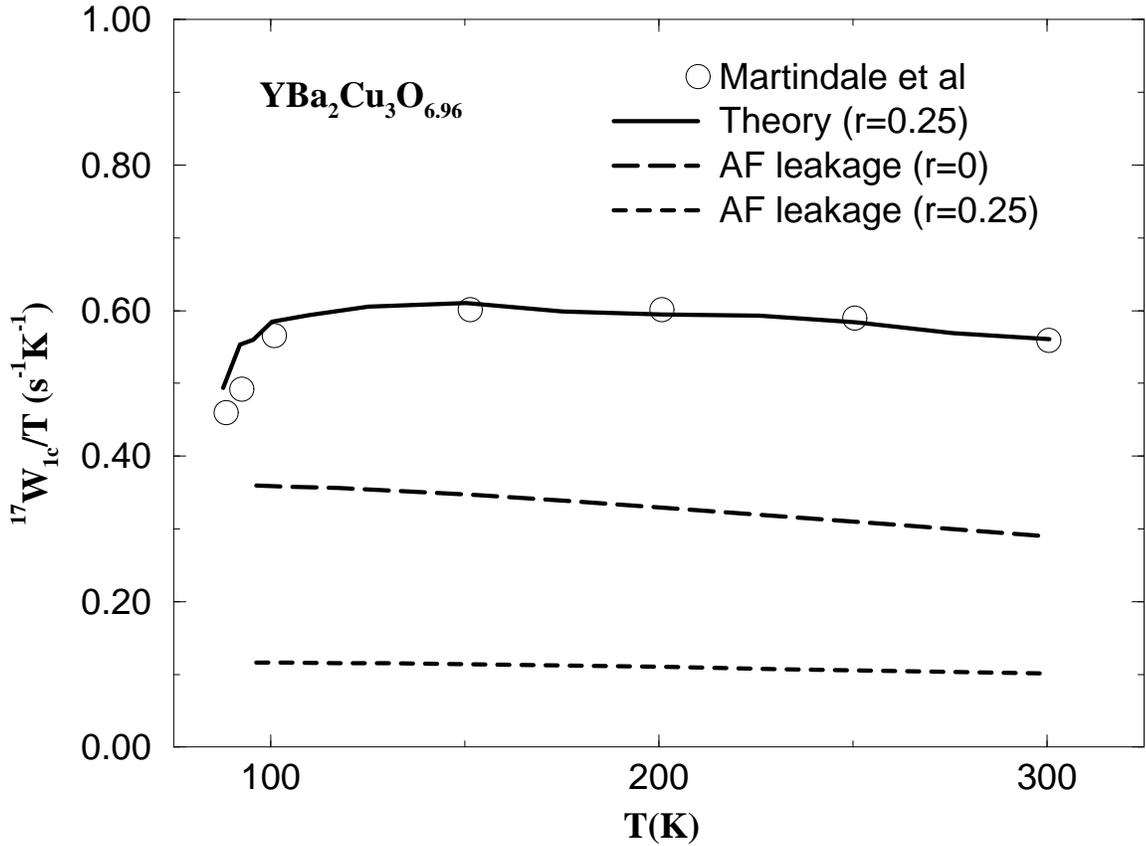}}\vspace{-2in}
\caption{The $\ox$ spin-lattice relaxation rate $^{17}W_{1c}/T$ 
calculated by assuming r=0.25 (solid line), plotted
against the experimental data of Martindale $et~al$\protect
\cite{martindale} (circles) for $\ybcoss$. Also shown is the 
contribution from AF leakage to the 
relaxation rates $^{17}W_{1c}/T$ calculated using our oxygen form factor 
with r=0.25 and the
standard Shastry-Mila-Rice form factor (r=0). 
}
\end{figure}

\begin{figure}
\centerline{\epsfxsize=15cm \epsfbox{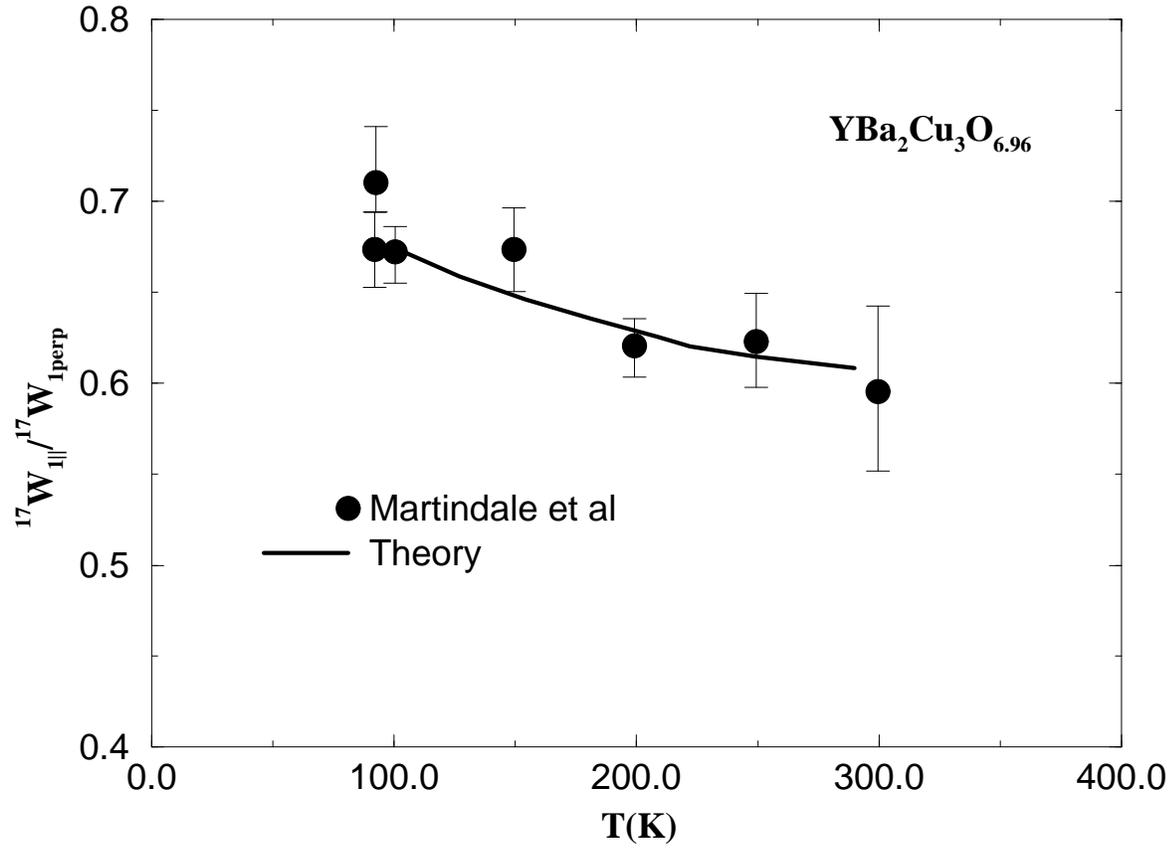}}\vspace{-2in}
\caption{ The temperature dependence of the oxygen relaxation rate ratios 
in $\ybcoss$ measured by Martindale $et~al$\protect\cite{martindale}, 
and fits using our theoretical expression Eq(\protect\ref{oxle}). 
}
\end{figure}

\begin{figure}
\centerline{\epsfxsize=15cm \epsfbox{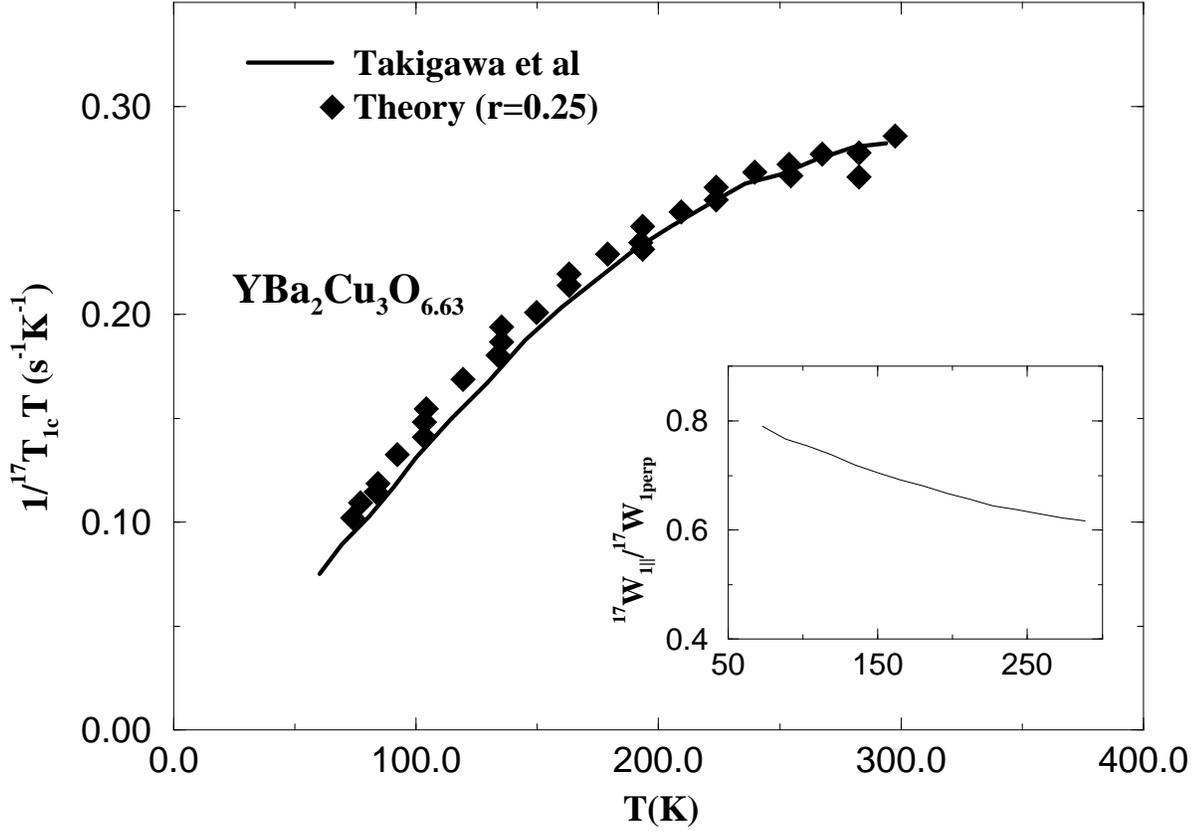}}\vspace{-2in}
\caption{The calculated spin-lattice relaxation rates $(^{17}T_{1c}T)^{-1}$ for
$\ybcox$,  compared with the data of 
Takigawa $et~al$.\protect\cite{takigawa} The inset shows the predicted 
anisotropy ratio as described in the text. The predicted anisotropy lies
slightly above the calculated curve for  $\ybcoss$ in Fig.8, yet both are
within the experimental error bars of the $\ybcoss$ 
data of Martindale $et~al$.\protect\cite{martindale}
}
\end{figure}
\end{document}